\documentclass[aps,prd,reprint,superscriptaddress,nofootinbib]{revtex4-2}

\usepackage{amsmath}
\usepackage{amssymb}
\usepackage{mathtools}
\usepackage{amsfonts}
\usepackage{graphicx}
\usepackage{xcolor}
\usepackage{epsfig}
\usepackage{color}
\usepackage{url}
\usepackage{comment}
\usepackage{bm}
\usepackage{mathrsfs}
\usepackage[utf8]{inputenc}
\usepackage{enumerate}
\usepackage{amsthm}
\usepackage{bbm}
\usepackage{upgreek}
\usepackage{soul}
\usepackage{esint}
\usepackage{tabularray}
\usepackage{multirow}
\usepackage[T1]{fontenc} 
\usepackage{orcidlink}

\usepackage{babel}

\graphicspath{{Plots/}}

\usepackage[capitalise]{cleveref}

\usepackage{verbatim}
\usepackage{hyperref}

\newcommand{\beq}{\begin{equation}}
\newcommand{\eeq}{\end{equation}}
\newcommand{\bea}{\begin{eqnarray}}
\newcommand{\eea}{\end{eqnarray}}
\newcommand{\bit}{\begin{itemize}}
\newcommand{\eit}{\end{itemize}}
\newcommand{\ben}{\begin{enumerate}}
\newcommand{\een}{\end{enumerate}}

\newcommand{\vinf}{v_{\infty}}

\def\scri{\mathscr{I}}
\def\hor{\mathscr{H}}

\newcommand{\STAG}{\affiliation{School of Mathematical Sciences and STAG Research Centre, University of Southampton, Southampton, SO17 1BJ, United Kingdom}}

\newcommand{\AEI}{\affiliation{Max Planck Institute for Gravitational Physics (Albert Einstein Institute), D-14476 Potsdam, Germany}}

\newcommand{\NBI}{\affiliation{Center of Gravity, Niels Bohr Institute, Blegdamsvej 17, 2100 Copenhagen, Denmark}}

\theoremstyle{plain}

\begin{document}
\title{Time-domain framework for the Teukolsky equation with a particle source using comoving hyperboloidal coordinates}

\author{Aditya Vaswani \orcidlink{0000-0003-4051-7461}} 
\STAG
\author{Leor Barack \orcidlink{0000-0003-4742-9413}} 
\STAG
\author{Oliver Long \orcidlink{0000-0002-3897-9272}}
\AEI
\author{Rodrigo Panosso Macedo \orcidlink{0000-0003-2942-5080}}
\NBI

\date{\today}

\begin{abstract}
We present a scheme and implementation code for time-domain integration of the Teukolsky equation in 1+1 dimensions with a point-particle source, based on comoving, spatially compactified hyperboloidal coordinates. We demonstrate that the scheme evades the problem of nonphysical growing modes that plague some numerical evolution schemes without compactification. Our use of comoving coordinates greatly simplifies the application of jump conditions on the particle's worldline. We develop our method and test its performance for a scalar field on a Schwarzschild background, first for a circular geodesic orbit source and then for a scattering geodesic orbit. We then present a test implementation of the method for the $s = -2$ Teukolsky equation, illustrating its long-term stability and absence of growing-mode behavior through comparison with similar results using noncompactified characteristic coordinates. Our method paves the way to calculations of the full gravitational self-force in extreme-mass-ratio scattering. 
\end{abstract}

\maketitle

\section{Introduction }

The problem of black-hole scattering dynamics has been attracting significant interest in recent years, motivated by advancements in observational capabilities of gravitational-wave campaigns. While astrophysical scattering events remain unlikely sources even for next-generation detectors, the precise modeling of their dynamics, especially in strong-field scenarios, can inform the calibration of gravitational-wave signal models for astrophysical, bound systems, for instance via the Effective One Body (EOB) infrastructure \cite{Damour2018, Damour2020, Damour2016}. Since the scattering process begins and ends with the two objects infinitely far apart, it admits well defined ``in'' and ``out'' states, making it amenable to techniques from effective-field theory and scattering amplitudes (see Refs. \cite{KalinPorto2020, KalinPorto_boundary2_2020, kosower_amplitudes_2019, Bern_etal_2021, Cheung_etal_2018} and references therein for a flavor of these developments). These techniques can leverage certain ``scatter-to-bound'' maps, which analytically continue some scattering observables to bound ones (e.g., scattering angle to periastron advance) \cite{KalinPorto2020, KalinPorto_boundary2_2020}.

One approach to black hole scattering is based on the gravitational self-force (GSF) formalism, tailored to the extreme-mass-ratio regime \cite{BarackPound2019, Pound2020}. In this approach, one systematically expands the Einstein field equations and the secondary's trajectory perturbatively in powers of the mass ratio $\mu/M$, assumed small. At leading order ($\mu/M \to 0$), the problem reduces to geodesic motion in the spacetime of the primary. Subleading terms then describe finite-$\mu$ corrections to the metric and the trajectory. A key advantage of the GSF approach is that it does not rely on a weak-field assumption: it is ``exact'' in the gravitational coupling $G$ and can thus be used reliably to study strong-field scenarios. In this way it is usefully complementary to weak-field perturbative approaches based on post-Minkowskian (PM) or post-Newtonian (PN) expansions.   

GSF methods have been applied extensively in studies of bound-orbit inspirals, but they are much less developed for scattering. Recent work by Warburton \cite{Warburton:2025GravRadScattering} (following early results in Refs. \cite{Hopper:2018Unbound, HoppperCardoso:2018Scattering}) presented numerical calculations of the energy loss in a scattering process through emission of gravitational waves, at leading-order in the mass ratio, and conducted a detailed comparison with PM results. A complete description of the scattering dynamics in the context of GSF theory, including conservative effects (which, for instance, impact the scattering angle) requires a full calculation of the local GSF acting on the particle along the scattering trajectory. Such a calculation remains a challenging task due to a variety of technical reasons (some described in some detail below), and it has not yet been presented.

Much of the development of GSF methodology for scattering has so far focused on a toy problem where the light black hole is replaced with a pointlike scalar charge, and one then considers the self-force exerted by the particle's own scalar field. This problem shares many technical similarities with the physical binary black hole problem, and thus forms a good testbed for method development. Reference~\cite{barack_self-force_2022} presented a numerical calculation of the scalar-field self-force along scattering orbits, with numerical results for the correction to the scattering angle due to both dissipation and conservative effects. This has prompted the derivation of a corresponding PM framework using Amplitudes methods, with results showing a good agreement with the numerical data \cite{barack_comparison_2023}. Enhanced numerical accuracy was achieved with the development of a frequency-domain self-force methodology for scattering \cite{Whittall:2023freqdom, Whittall_etal:2025Gegenbauer} and a subsequent time/frequency-domain hybrid \cite{Long:2024ltn}. This hybrid framework was used in Ref.~\cite{Long:2024ltn} to demonstrate how strong-field GSF information can be used to construct resummed PM models with uniform accuracy across the whole of the scattering parameter space.

The GSF methods developed for the scalar-charge problem involve numerical integration of the Klein-Gordon field equation on the background of the primary black hole (assumed to be Schwarzschild in all work so far). To apply these methods to the physical black hole problem, one instead needs to solve the appropriate metric perturbation equations. One approach, which most directly extends the scalar-field problem to gravity, is based on the Teukolsky formulation of black hole perturbation theory. This approach was taken in Ref.~\cite{Long:2021TDmetric}, in which a scheme was developed for the numerical integration of the spin $s=\pm 2$ sourced Teukolsky equation on a Schwarzschild background in the time domain, decomposing in spin-weighted spherical harmonics and tackling the modal field equations in 1+1 dimensions (time+radius). The integration scheme was based on a straightforward finite-difference formula in characteristic (Eddington-Finkelstein) coordinates, starting from characteristic initial data. These initial data were set to zero, with the resulting burst of transient ``junk'' radiation later discarded. A similar scheme has been used extensively over the years in GSF work for a direct integration of the Lorenz-gauge metric perturbation equations \cite{BarackSago:2007:GSF_circ, BarackSago:2010:GSF_Ecc}, and it is also the scheme upon which the scattering calculation of Ref.~\cite{barack_self-force_2022} is based. 

The analysis of Ref.~\cite{Long:2021TDmetric} identified a problem with the application of the above scheme to the $s=\pm 2$ Teukolsky equation. In contrast to the situation in the scalar-field ($s=0$) case, here there exist nonphysical, advanced-boundary-conditions solutions of the equation that grow unbounded in time. For $s=-2$, these solutions grow along future null infinity as $\sim u^4$, and for $s=+2$, they grow along the future event horizon as $\sim\exp(v/2M)$ (here $u$ and $v$ are the retarded and advanced Eddington-Finkelstein time coordinates). Since in a characteristic evolution one cannot (and need not) impose boundary conditions, one has \textit{a priori} no way in which to eliminate such unphysical growing modes. Indeed, the experiments of Ref.~\cite{Long:2021TDmetric} revealed that such modes, seeded from junk initial data and numerical roundoff, always grow to dominate numerical solutions, rendering them unusable. A similar behavior was observed by others, in characteristic numerical solvers for the $s=\pm 2$ Green's function \cite{OToole:2021RWZ, OToole:thesis}, and also in some Cauchy evolution schemes \cite{Vishal2:pers_comm}. In time-domain treatments with causal boundaries, such modes may be controlled via boundary conditions  \cite{Bishoyi:2026dwl}.  

In Ref.~\cite{Long:2021TDmetric}, the problem of growing modes was circumvented by working with the Regge-Wheeler equation instead of the Teukolsky equation. However, in that approach the reconstruction of the metric perturbation in a gauge suitable for GSF calculations is then numerically inefficient (as it necessitates taking a large number of numerical derivatives), and, moreover, it is not generalizable to the case of a Kerr primary.  A Teukolsky-based approach is greatly advantageous computationally, but one must find a way around the growing-mode problem.

It was proposed in \cite{Long:2021TDmetric,barack_self-force_2022} that a remedy might be found in the use of compactified hyperboloidal coordinates. In such schemes, retarded boundary conditions are automatically satisfied, suggesting the schemes are immune to the growing-mode problem. Intuitively, the use of spatially compactified hyperboloidal time slices that intersect future null infinity and the future horizon should ensure that any unphysical (advanced) radiation component is infinitely under-resolved in these two asymptotic regimes. 

Methods based on hyperboloidal foliation with spatial compactification have become ubiquitous in recent years: see, for example, Refs.~\cite{Zenginoglu:2008uc,Harms:2013ib,PanossoMacedo:2014dnr,Csukas:2019kcb,Csukas:2021sia} for applications to the vacuum Teukolsky equation and Refs.~\cite{Harms:2014dqa,OBoyle:2023jqo,DaSilva:2023xif,Vishal:2023fye,DaSilva:2024yea,Vishal:2025pqc,Roy:2025kra} for time-domain implementations with a point particle source. Hyperboloidal foliation methods have also been applied in frequency-domain treatments, as in Refs.~\cite{PanossoMacedo_etal:2022HypFreqSF,Leather:2024GSF_HypSpectral,Roy:2025kra}. 
We could not find mention of the growing-mode problem in the above time-domain literature, which supports our suspicion that compactification averts the issue. However, to the best of our knowledge, there has not been a direct demonstration that this is true. Moreover, there has been some indication that growing-mode behavior can persist even in schemes with compactification
\cite{Vishal2:pers_comm}.

In this work, we develop a new time-domain integrator for the sourced $s=-2$ Teukolsky equation based on a new choice of compactified hyperboloidal coordinates, and demonstrate that, indeed, the scheme is not susceptible to growing modes. A key element of our architecture is the use of source-adapted, {\em comoving} coordinates, which greatly simplifies the application of jump conditions along the worldline of the particle sourcing the perturbation. (An alternative approach, utilizing comoving double-null coordinates was recently developed by Roy {\it et al.~}in Ref.~\cite{Roy:2025kra} for the study of plunging scalar-charge trajectories. General geometrical aspects of source-adapted hyperboloidal foliations were explored in Ref.~\cite{Zenginoglu:2025_BridgingTimes}.) We first develop and implement our method for the well-studied case of a scalar field, where comparison numerical results are in abundance. We show a good agreement with existing numerical results for both bound and scattering orbits. Then, we apply our method to the $s=-2$ Teukolsky equation satisfied by the Hertz potential used for generating metric perturbations in the so-called ``no-string'' radiation gauge \cite{Pound:2013faa}. We demonstrate that the numerical evolution proceeds stably for a long time without a hint of interference from the growing modes, and check our results against analytical expressions and numerical results obtained using other methods. We directly contrast these findings with results from the code of Ref.~\cite{Long:2021TDmetric} using uncompactified characteristic coordinates.

The structure of this paper is as follows. In Sec.~\ref{sec:scattering_Schwarzschild}, we briefly review scattering geodesics in Schwarzschild spacetime, providing essential results and notation for later sections. In Sec.~\ref{sec:hyperboloidal_comoving}, we introduce our comoving hyperboloidal coordinates on Schwarzschild spacetime and discuss their pertinent properties. Section \ref{sec:scalar_field_hyper} obtains the vacuum Klein-Gordon equation in the new coordinates and formulates jump conditions along the trajectory of a sourcing scalar charge, ready for numerical implementation. Our finite-difference scheme is described in Sec.~\ref{sec:fin_diff}. Section \ref{sec:num_tests} then presents various numerical tests and comparisons with characteristic-evolution results from the code of Ref.~\cite{barack_self-force_2022}. In Sec.~\ref{sec:hertz_Teuk}, we finally apply our method to the $s=-2$ Teukolsky equation and illustrate the resolution of the growing-mode problem through comparison with similar results from the characteristic code. Section~\ref{sec:discussion} discusses our results and the next steps toward a full self-force calculation. In the Appendix, we present analytical expressions for the jumps in the Hertz potential modes and their derivatives, necessary as input for our numerical implementation in Sec.~\ref{sec:hertz_Teuk}.

Throughout this work, we use geometrized units, with $G = 1 = c$, and metric signature $({-}{+}{+}{+})$.

\section{Review of geodesic scattering in Schwarzschild spacetime}
\label{sec:scattering_Schwarzschild}

We shall develop our new comoving hyperboloidal scheme in the next section  primarily with a mind to implement it for scattering-orbit sources. For that reason, it is useful to start with a brief review of scattering geodesics in Schwarzschild spacetime, introducing relevant notation and several results that will be of use later.   

In our setup, the primary object is assumed to be a Schwarzschild black hole with mass parameter $M$ and line element 
\beq
\label{eq:SchwarzschildMetric}
ds^2 = -f(r) ( dt^2 - dr_*^2) + r^2 d\Omega^2.
\eeq
Here, $f(r) \coloneqq 1 - 2M/r$, $d\Omega^2 = d\theta^2 + \sin\theta^2\,d\varphi^2$, and the tortoise radial coordinate
\begin{equation}
\label{r*}
    r_* = r + 2M \ln[r/(2M)-1]
\end{equation}
satisfies $dr_*/dr=1/f$.
We let $y^\alpha = y^\alpha_p (\uptau)$ define the geodesic trajectory of a test point particle in the Schwarzschild geometry, parametrized by proper time $\uptau$, with 4-velocity $u^\alpha \coloneqq d y_p^\alpha / d \uptau$. Without loss of generality we restrict the motion to the equatorial plane, such that $\theta_p \equiv \pi/2$ and $u^\theta \equiv 0$. The geodesic admits conserved (specific) energy and angular momentum 
\bea \label{E&L}
E \coloneqq f_p u^t  \quad \text{and} \quad
L \coloneqq r_p^2 u^\varphi,
\eea
where $f_p \coloneqq f(r_p)$. The particle's radial motion is governed by
\begin{align}
u^r = \pm \sqrt{E^2-V(r_p;L)},
\label{rdot}
\end{align}
where the negative (positive) sign corresponds to the inbound (outbound) leg of the geodesic, and
\bea
V(r;L) = f\left(1+\frac{L^2}{r^2}\right)
\eea
is the effective potential around the black hole. 

A (timelike) scattering geodesic in this spacetime is a member of a 2-parameter family of curves, usually parametrized in terms of $(E,L)$, or in terms of an eccentricity $e$ and semilatus rectum $p$ defined as the relevant roots of 

    \begin{align}
        E^2 = \frac{(p-2)^2-4e^2}{p(p-3-e^2)}, \qquad L^2 = \frac{p^2M^2}{p-3-e^2}.
    \end{align}

Alternative common parameters in the scattering case are the 3-velocity magnitude at infinity, $v_\infty$, and the impact parameter, $b$, defined as
    \begin{subequations}
    \begin{align}
        \vinf &\coloneqq \lim_{\uptau \to - \infty} \sqrt{\frac{( u^r )^2 + ( r_p u^\varphi )^2}{(u^t)^2}} = \frac{\sqrt{E^2 - 1}}{E} 
        \\
        b &\coloneqq \lim_{\uptau\to-\infty} r_p(\uptau)\sin\left|\varphi_p(\uptau)-\varphi_p(-\infty)\right|\ = \frac{L}{v_\infty E}.
    \end{align}
    \end{subequations}

Scattering geodesics satisfy 
\begin{align}
    E > 1, \quad L > L_c(E),
\end{align}
\begin{align}
    e > 1, \quad p > 6+ 2e,
\end{align}
or 
\begin{align}
    0< \vinf <1, \quad b > b_c(v) ,
\end{align}
where (see, e.g., Ref.~\cite{barack_self-force_2022})
\begin{align}
L_c(E):= M \sqrt{\frac{9 E^2 (3 E^2 - 4) + E (9 E^2 - 8)^{3/2} + 8}{2 (E^2 - 1)}}
\end{align}
and $b_c(\vinf):=L_c(E)/\sqrt{E^2-1}$, with $E=(1-\vinf^2)^{-1/2}$.

In terms of the $(e, p)$ parametrization, the radial and azimuthal motions are described by
\begin{align}
    r_p (\chi) = \frac{M p}{1 + e \cos{\chi}}
\end{align}
and
\begin{align}\label{varphi_p}
    \varphi_p (\chi) =  k \sqrt{p/e}\, \, \text{El}_1 \bigg ( \frac{\chi}{2}; -k^2 \bigg ),
\end{align}
where $k^2 = 4 e/(p - 6 - 2 e)$ and $\text{El}_1$ is the incomplete Elliptic integral of the first kind.
The parameter $\chi$ is a relativistic anomaly taking values in $- \chi_{\infty} < \chi < \chi_{\infty}$, with $\chi_{\infty} = \arccos{(-1/e)}$ and periapsis at $\chi = 0$. In Eq.~(\ref{varphi_p}), we have chosen the initial condition $\varphi_p(-\chi_\infty)=0$. The total scattering angle is given by 
\begin{align}
    \Delta \varphi_p &= \varphi_p (\chi_{\infty}) - \varphi_p (-\chi_{\infty}) - \pi.
\end{align}

\begin{figure}[htb]
    \centering
    \includegraphics[width=0.5\textwidth]{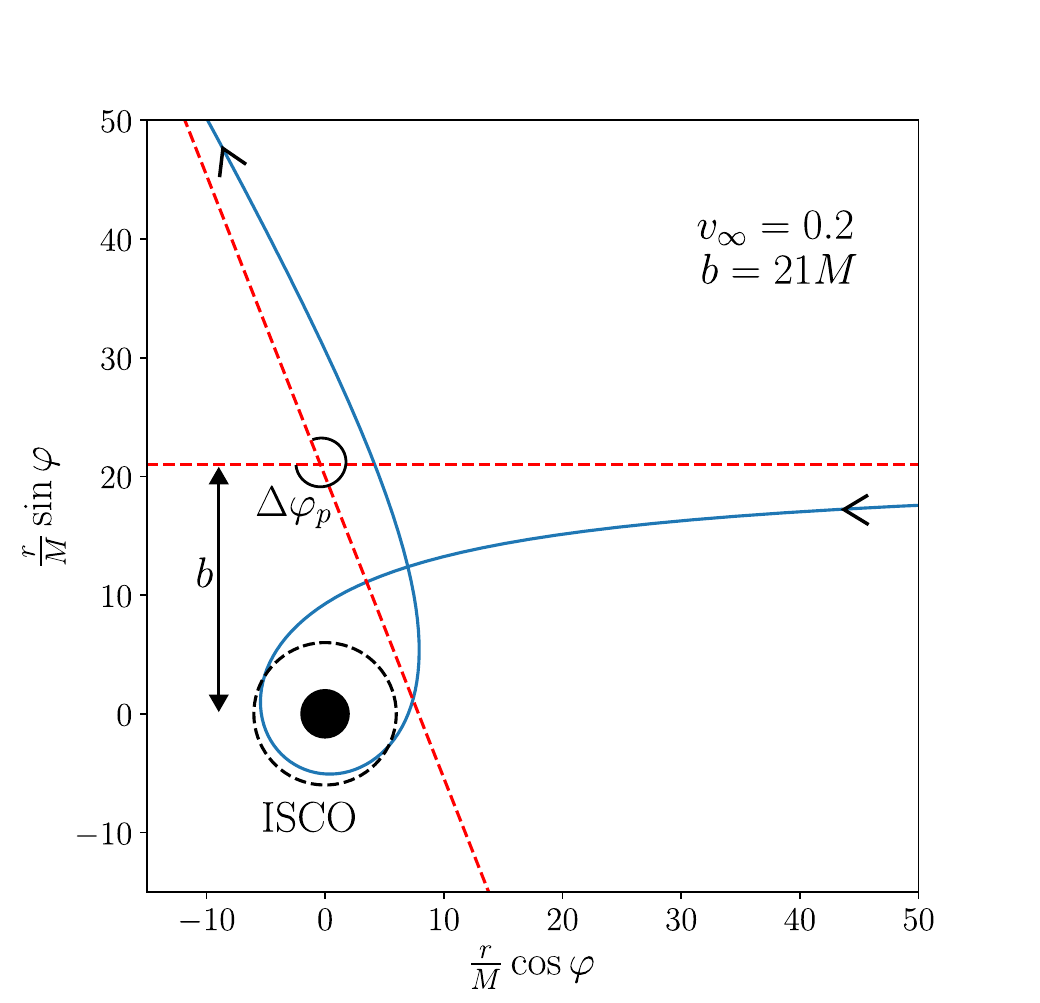}
    \caption{A sample scattering orbit with $(v_\infty, b) = (0.2, 21M)$. The black disc represents the central Schwarzschild black hole of mass $M$, with the Innermost Stable Circular Orbit (ISCO) indicated in black dashed line for reference.  The dashed red lines are asymptotic to the scattering geodesic at $t\to\pm\infty$, with $\Delta \varphi_p\approx 301^{\circ}$ being the scattering angle. This will be our reference geodesic for various tests in this work. }
    \label{fig:scatter_orbit}
\end{figure}

A section of a sample scattering orbit with $(\vinf,b)=(0.2,21 M)$ is depicted in Fig~\ref{fig:scatter_orbit}. This will be our reference scattering geodesic for various numerical tests in this work.  

\section{Comoving Hyperboloidal Coordinates} \label{sec:hyperboloidal_comoving}

In this section, we present a comoving hyperboloidal coordinate system adapted to a fixed timelike worldline $y^\alpha_p (\uptau)$, assumed to be a (nonplunging) geodesic of the Schwarzschild background. In Schwarzschild (tortoise) coordinates, the worldline is described by 
\begin{align}
(t,r_*,\theta,\varphi)=(t_p(\uptau),M x_p(\uptau),\pi/2,\varphi_p(\uptau)),
\end{align}
where, recall, $\uptau$ is proper time along the worldline.

We define our hyperboloidal coordinates through a transformation $(t, r^*, \theta, \varphi) \to (\tau, \sigma, \theta, \varphi)$, where $\tau$ is a (dimensionless) time coordinate (not to be confused with the proper time $\uptau$) such that $\tau={\rm const}$ hypersurfaces are hyperboloidal~\cite{Zenginoglu:2007jw,PanossoMacedo:2023qzp}. The coordinate $\sigma$ is a compactified radial coordinate that decreases monotonically with $r$ from $\sigma=1$ on the future event horizon ($\hor^+$) to $\sigma=0$ at future null infinity ($\scri^+$). Hence, the exterior of the black hole is covered by
\begin{equation}
    0\leq \sigma \leq 1, \quad \quad -\infty < \tau < \infty\, .
\end{equation}
Specifically, $\tau$ and $\sigma$ are defined via
\begin{subequations}\label{trans}
\begin{align}
\label{t_trans}
t/M &= \tau + H(\sigma),
\\ 
\label{r*_trans}
r_*/M &= \alpha(\sigma) x_p(\tau) + K(\sigma),
\end{align}
\end{subequations}
with the ``height'' and ``shift'' functions 
\begin{subequations}
\begin{align}
\label{H}
H (\sigma) &= \frac{1}{\sigma^2} - 2\ln[\sigma(1-\sigma)] ,
\\
\label{K}
K (\sigma) &= \frac{1}{\sigma^2} + 2\ln[(1-\sigma)/\sigma] - 4. 
\end{align}
\end{subequations}
The purpose of the function $\alpha(\sigma)$ is to suppress the support of the term $x_p(\tau)$ near the event horizon, the need for which will be explained further below. A specific form for $\alpha(\sigma)$ will also be prescribed below, but for the time being it suffices to define it as any differentiable function on $0\leq \sigma\leq 1$ with the properties
\begin{align}
    \alpha'(\sigma)\leq 0, \quad \alpha(1)=0, \quad \alpha(\sigma\leq 1/2)\equiv 1, 
\end{align}
where a prime denotes $d/d\sigma$. Note that
\begin{align}
    \sigma=1/2\quad\Leftrightarrow\quad r_* =Mx_{p}(\uptau),
\end{align}
which means that the particle remains at the fixed spatial coordinate value of $\sigma=1/2$ at all time; hence the labeling of our coordinates as ``comoving''.

In terms of the new coordinates, the line element is
\begin{align}
ds^2 =& M^2 f 
\big[ 
-\gamma d\tau^2 
+2(\alpha \dot{x}_p \beta - \eta) d\tau d\sigma  
\nonumber \\
&\quad + (\beta^2 -\eta^2) d \sigma^2 \big] + r^2 d \Omega^2,
\end{align}
where
\begin{subequations}\label{gamma_beta_eta}
\begin{align}
\gamma (\tau,\sigma) &\coloneqq 1 - (\alpha \dot{x}_p)^2, 
\\
\beta(\tau,\sigma) &\coloneqq K' + \alpha' x_p,
\\
\eta(\sigma) &\coloneqq H'.
\end{align}
\end{subequations}
Hereafter, we use a prime to denote $\partial_\sigma$ (at fixed $\tau$, when acting on functions of both $\sigma$ and $\tau$) and an overdot to denote $\partial_\tau$ (at fixed $\sigma$). In what follows, we check and confirm that $(\tau,\sigma,\theta,\varphi)$ form a legitimate set of hyperboloidal coordinates on the black hole's exterior. 

First, we check that each $\tau={\rm const}$ hypersurface is  everywhere spatial, as required. This requires $g^{\tau\tau}<0$ for all $\tau$, all $0<\sigma<1$, and all $r_p(\tau)$ corresponding to a (nonplunging) timelike geodesic. We have
\beq\label{gup_tautau}
g^{\tau\tau} = fM^2(\beta^2-\eta^2)/\Delta,
\eeq
where $\Delta$ is the determinant of the time-radial part of the metric, the sign of which is invariant under regular coordinate transformations. Hence $\Delta<0$, which can also be checked with a direct calculation:  
\begin{eqnarray}\label{Delta}
\Delta = -f^2 M^4 \left(\beta - \alpha \dot{x}_p \eta\right)^2 <0 .
\end{eqnarray}
Therefore the sign of $g^{\tau\tau}$ is that of 
\begin{align}
    \label{g^tautau proof}
    \eta^2 - \beta^2 = - \frac{16 ( 1 + \sigma^2 ) - 4 (1 - \sigma + \sigma^2) x_p \alpha'}{\sigma^3 ( 1 - \sigma )}
    - ( \alpha' x_p )^2,
\end{align}
which is manifestly negative definite, noting $1 - \sigma + \sigma^2>0$, $\alpha'\leq 0$, and the fact that for a nonplunging timelike geodesic we always have $x_p>x_p(r=3M)\simeq 1.6137>0$. 
Hence $g^{\tau\tau}<0$, confirming that $\tau$ is a legitimate time coordinate. 

Second, let us check that $\sigma$ is indeed monotonically decreasing with $r_*$, and hence a legitimate spatial coordinate. From (\ref{r*_trans}) we find
\begin{align}
r_*'/M =\alpha'(\sigma)x_p  -\frac{2}{\sigma^3} - \frac{2}{\sigma ( 1 - \sigma )} ,
\end{align}
where we again recall $\alpha'(\sigma)\leq 0$ and $x_p>0$, and observe the last two terms are manifestly negative definite. Therefore $r_*'<0$ for all $\tau$ and all $0<\sigma<1$, as required.

Third, we check that $\tau={\rm const}$ hypersurfaces are indeed ``hyperboloidal''. Introducing the Eddington-Finkelstein null coordinates $v:=t+r_*$ and $u:=t-r_*$ and recalling Eq.~(\ref{trans}), we find, for each $\tau={\rm const}$ hypersurface,  $\sigma\to 0 \Rightarrow u\to {\rm const}$ and $v\to\infty$, showing the hypersurface intersects $\scri^+$.  Similarly, on each $\tau={\rm const}$ hypersurface, $\sigma\to 1 \Rightarrow v\to {\rm const}$ and $u\to\infty$, showing the hypersurface intersects $\hor^+$. This confirms that our coordinates indeed define a hyperboloidal foliation of the black hole's exterior. It can be checked that our $\tau={\rm const}$ hypersurfaces are asymptotically null at $\scri^+$, but remain spatial at $\hor^+$.

Finally, we need to check that all worldlines $\sigma={\rm const}$ (with constant $\theta,\varphi$) are everywhere {\em timelike} (except the ones with $\sigma=0$ or $\sigma=1$, which are, of course, null). This requirement ensures that our Cauchy evolution scheme does not propagate data superluminally. In terms of the metric, the condition is
\begin{equation}\label{g_tautau}
g_{\tau\tau}=-M^2 f(1-\alpha^2\dot{x}_p^2)<0
\end{equation}
for all $\tau$ and all $0<\sigma<1$. To see this is true, recall $0\leq\alpha\leq 1$, and note  
\begin{align}\label{xdot}
    \dot{x}_p^2 =f_p^{-2}\left(\frac{dr_p}{dt}\right)^2 = (u^r/E)^2 =  1 - \frac{V}{E^2} <1,
\end{align}
where in the third equality we have used the radial geodesic equation (\ref{rdot}). Hence in Eq.~(\ref{g_tautau}) we have $\alpha^2 \dot{x}_p^2<1$, confirming $g_{\tau\tau}<0$ as desired.

The purpose of the factor $\alpha(\sigma)$ in Eq.~\eqref{r*_trans} is to suppress the contribution of the $x_p$ term near $\hor^+$, in order to let the behavior there be controlled by the $K(\sigma)$ term rather than the $x_p(\tau)$ term. To see why this is necessary, consider the asymptotic form of the transformation near $\hor^+$ on each $\tau=\text{const}$ slice,
\begin{align}
\label{sig_scri_H}
\sigma \simeq 1 - f(r) \, \exp\left[-\alpha(1) x_p(\tau)/2 + 5/2\right] ,
\end{align}
obtained from Eq.~\eqref{r*_trans} using the asymptotic relations $r_*/M \sim 2 + 2 \ln f$ and $K(\sigma) \sim 2 \ln(1 - \sigma) - 3$ as $\sigma \to 1$.
Note how, had we not set $\alpha(1)=0$, the difference $1-\sigma$ would have been exponentially suppressed at large $x_p(\tau)$. In the case of scattering this would have led to a catastrophic loss of numerical resolution near the horizon at early and late times, where $x_p$ grows unbounded: to sample the near-horizon region would have required impractically fine stepping in $\sigma$. 

In addition to $\alpha(1)=0$ we must require $\alpha(1/2)=1$ to ensure $r_*(\sigma=1/2)=Mx_{p}$, and in the above checks we have also had to assume $\alpha'(\sigma)\leq 0$ as well as $|\alpha|\leq 1$ for all $0\leq\sigma\leq 1$.  Unfortunately, these conditions cannot be all satisfied simultaneously by any analytical function $\alpha(\sigma)$. We therefore allow $\alpha$ to have a mild non-smoothness, which for convenience we choose to restrict to the particle's location at $\sigma=1/2$. Specifically, we choose
\begin{align}
\label{alpha1}
\alpha(\sigma) &= \left\{ \begin{array}{ll}
        1-(2\sigma-1)^{N+1},  & \sigma>1/2, \\
        1 , &     \sigma\leq 1/2 ,
    \end{array} \right.
\end{align}
with $N$ being an integer. This function has a $C^N$ regularity at the particle (and is smooth elsewhere). In our particular numerical implementation for this work, we take $N=4$, sufficient to ensure that no spurious distributional terms occur in our finite-difference representation of the field equation. Different values of $N$ may be motivated depending on the particular scheme used.

We note our coordinates' property $\sigma\sim 1/r^2$ near $\scri^+$, which differs from the more conventional choice $\sigma\sim 1/r$ \cite{Ansorg:2016ztf, PanossoMacedo:2018hab, PanossoMacedo:2019npm}. Our choice ensures a better sampling of the wave-zone region at large $|\tau|$ (i.e., in the early and late stages of the scattering), when the $x_p$ term in Eq.~(\ref{r*_trans}) is large. Alternative choices with $\sigma\sim 1/r^{n>2}$ may be motivated in future work, depending on the maximal value of $x_p$ in the simulation; see the discussion at the end of Sec.~\ref{subsec:scattering}.

\begin{figure}[h!]
    \centering
    \includegraphics[width=\linewidth]{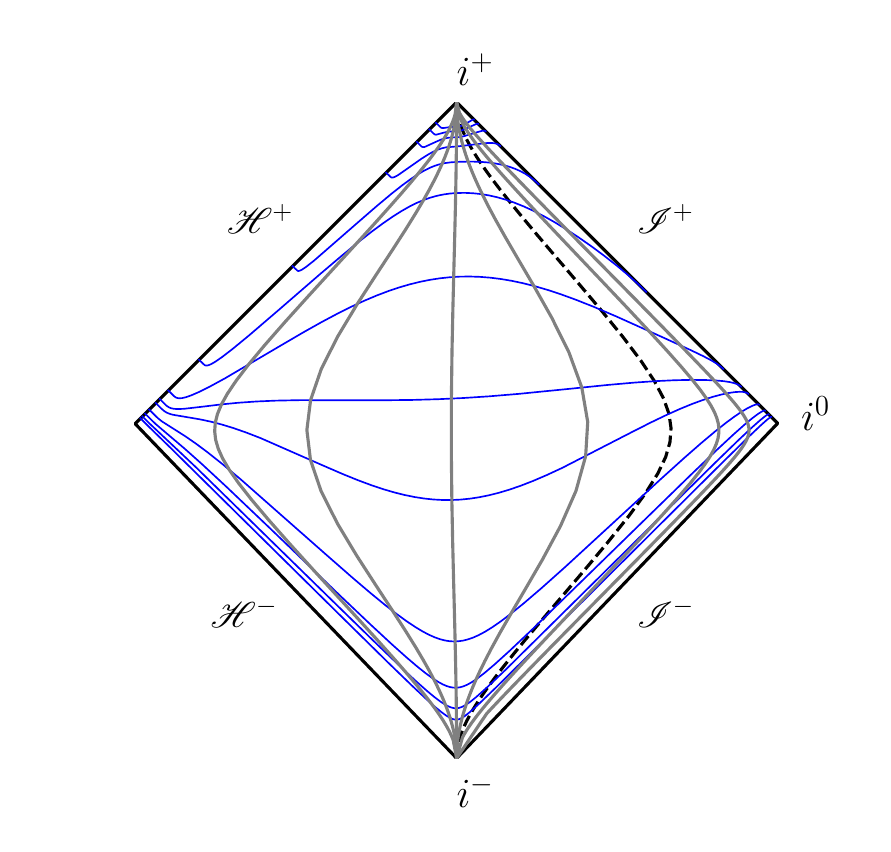}
    \caption{An illustration of our hyperboloidal coordinate grid, for the case $r_p(\tau)={\rm const}=4M$, showing a sample set of $\tau={\rm const}$ curves (blue) and $\sigma={\rm const}$ curves (grey) on a conformal diagram covering the exterior of the black hole. The comoving curve $\sigma=1/2$ is shown as a thick dashed black line. To plot this diagram, we have conformally mapped the coordinates $t(\tau,\sigma)$ and $r_*(\tau,\sigma)$ to the 2D plane with Cartesian coordinates $(x,y)$ using $y= \pi^{-1} (\arctan{(v/2)} + \arctan{(u/2)})$ and $x= \pi^{-1} (\arctan{(v/2)} - \arctan{(u/2)})$, where $v=t+r_*$ and $u=t-r_*$. }
    \label{fig:penrose_diag}
\end{figure}

Figure \ref{fig:penrose_diag} illustrates our coordinate grid using a conformal diagram of the black hole's exterior. For this illustration we use a fiducial particle worldline with $r_p(\tau)={\rm const}=4M$.

\section{Scalar field equation and multipole decomposition}
\label{sec:scalar_field_hyper}

Our goal is to numerically solve the scalar field equation sourced by a pointlike scalar charge on a fixed geodesic orbit outside the black hole. We will do this using the comoving hyperboloidal coordinates introduced in the previous section, with the timelike worldline $y^\alpha_p (\uptau)$ taken to be that of the scalar charge. We take the particle's charge to be $q$, and assume it sources a scalar field $\Phi$ that satisfies the massless, minimally-coupled Klein-Gordon equation (or, equivalently, the Teukolsky equation with $s=0$ and $a=0$),
\begin{align} \label{eq:Klein-Gordon}
g^{\alpha\beta}\nabla_{\alpha}\nabla_{\beta}\Phi&= 
\frac{1}{\sqrt{-g}} \partial_{\mu} \big ( \sqrt{-g} g^{\mu \nu} \partial_{\nu} \Phi \big ) \nonumber\\
&= -4 \pi q \int \frac{\delta^4 ( x^{\alpha} - y^{\alpha}_p (\uptau))}{\sqrt{-g}} d\uptau .
\end{align}
Here, $g_{\alpha\beta}$ is the background Schwarzschild metric, $\nabla_{\beta}$ is the covariant derivative associated with it, and $g$ is its determinant. We decompose the field into spherical harmonic modes in the form 
\beq\label{Ylm_decomposition}
\Phi(t,r,\theta,\varphi) = \dfrac{q}{r}\sum_{\ell m} \phi_{\ell m} (t,r) \, Y_{\ell m}(\theta, \varphi)\, .
\eeq
Each time-radial mode satisfies the wave equation 
\beq \label{eq:DecomposedKG}
\square_{\ell m} \phi_{\ell m} =  S_{\ell m}(\uptau)\delta(r-r_p(\uptau))\, ,
\eeq
where
\beq
\square_{\ell m} :=\frac{\partial^2}{\partial t^2} - \frac{\partial^2}{\partial r_*^2} + \tilde V_\ell(r)/M^2,
\eeq
and
\beq
S_{\ell m}(\uptau) = \frac{4 \pi f_p^2}{E r_p}{Y}^*_{\ell m}(\pi/2,\varphi_p),\label{eq:source}
\eeq
where an asterisk denotes complex conjugation, and where 
\beq
\tilde V_\ell(r) = M^2 f(r) \left( \frac{\ell(\ell+1)}{r^2}+\frac{2M}{r^3} \right)
\eeq
is an (a-dimensionalized) potential. 

Transforming to $\tau,\sigma$ coordinates using Eqs.~(\ref{trans}), the modal equations (\ref{eq:DecomposedKG}) take the form 
\begin{align}
    &\ddot{\phi}_{\ell m} + C_{\tau \sigma} \dot{\phi}'_{\ell m} + C_{\sigma \sigma} \phi''_{\ell m} +C_{\tau} \dot{\phi}_{\ell m}
    \nonumber \\
    &\quad  + C_{\sigma} \phi'_{\ell m} + C_0 \phi_{\ell m} = \tilde S_{\ell m}(\tau)\delta\left(\sigma-1/2\right),
    \label{modal_KG_eqn_hyperboloidal}
\end{align}
where the $C$ coefficients are functions of $\sigma$ as well as $\tau$ [via $x_p(\tau)$], and we recall an overdot and prime denote $\partial_\tau$ and $\partial_\sigma$, respectively.  Explicitly, we find
\begin{subequations}
\label{KG_coeffs_compact}
\begin{align}
    C_{\tau \sigma} &= 2 \frac{\eta - \alpha \beta \dot{x}_p}{\beta^2 - \eta^2},
    \\
    C_{\sigma \sigma} &= \frac{- \gamma}{\beta^2 - \eta^2},
    \\
    C_{\tau} &= \frac{\alpha \eta}{\zeta} \ddot{x}_p + \frac{\gamma ( \beta \eta' - \beta' \eta ) + \eta ( \beta \gamma' + 2 \eta \dot{\beta} ) }{\zeta ( \beta^2 - \eta^2 )},
    \\
    C_{\sigma} &= \frac{- \alpha}{\zeta} \ddot{x}_p + \frac{\gamma ( \beta' - \alpha \eta' \dot{x}_p ) - \beta \gamma' - 2 \eta \dot{\beta} }{\zeta ( \beta^2 - \eta^2 )},
    \\
    C_0 &= \frac{\zeta^2}{\beta^2 - \eta^2} \tilde V_\ell(r),
\end{align}
\end{subequations}
where
\begin{equation}
\zeta (\tau, \sigma) := \beta (\tau, \sigma) - \alpha (\sigma) \eta (\sigma) \dot{x}_p (\tau);
\end{equation}
recall the definitions of $\gamma$, $\beta$, and $\eta$ from Eqs.~(\ref{gamma_beta_eta}).
The source function in Eq.\ (\ref{modal_KG_eqn_hyperboloidal}) reads
\begin{align}
\label{tildeS}
    \tilde S_{\ell m}(\tau) = \frac{\pi M f_p (3 - 2 \dot{x}_p)}{10 E r_p}{Y}^*_{\ell m}(\pi/2,\varphi_p).
\end{align}

\subsection{Jump conditions at the particle}
\label{subsec:jumps}

Since the inhomogeneous differential equation \eqref{modal_KG_eqn_hyperboloidal} is sourced by a delta function, we expect solutions to be generally non-differentiable across the particle's worldline at $\sigma=1/2$. The jumps in the field's derivatives can be readily derived by inserting the ansatz
\begin{align}
\phi (\tau, \sigma) = \phi^> (\tau, \sigma) \Theta(\sigma - 1/2) 
+ \phi^< (\tau, \sigma) \Theta(1/2 - \sigma) 
\label{phi_ansatz}
\end{align}
(with $\Theta$ being the Heaviside step function) into the field equation \eqref{modal_KG_eqn_hyperboloidal}, and demanding that it is satisfied distributionally. (Throughout this discussion, we drop the $\ell m$ labels for brevity.) This procedure reveals that $\phi^{\lessgtr}$ are homogeneous solutions, and that the field is continuous at the particle:
\begin{align}\label{J=0}
J(\tau) :=\: \left( \phi^> - \phi^< \right) \Big|_{\sigma=1/2} = 0. 
\end{align}
We also find
\begin{align}
C_{\tau \sigma} J_{\tau} + C_{\sigma \sigma} J_{\sigma} = \tilde{S}, 
\end{align}
where we have defined $J_\alpha(\tau) \coloneqq \left( \phi^>_{,\alpha} - \phi^<_{,\alpha}\right)\Big|_{\sigma=1/2}$, 
and the $C$ coefficients are evaluated at $\sigma=1/2$. Thanks to the fact that $\tau,\sigma$ are comoving coordinates, we have $J_\tau = dJ/d\tau$, which, by virtue of (\ref{J=0}), implies
\begin{align}
J_{\tau} = 0 .
\end{align}
This, in turn, gives
\begin{align}
\label{Js}
    J_\sigma =\left.\frac{\tilde S}{C_{\sigma\sigma}}\right|_{\sigma=\frac{1}{2}}  = -\frac{32\pi M f_p}{Er_p}\frac{(3 - 2 \dot{x}_p)}{1 - \dot{x}_p^2} Y^*_{\ell m}(\pi/2, \varphi_p).
\end{align}

We will make direct use of the above jump conditions in our finite-difference numerical integration scheme, to be presented in the next section. As we shall see, we will further need expressions for the jumps in the second, third, and fourth derivatives of the field across the particle. These we obtain sequentially with the help of the field equation (\ref{modal_KG_eqn_hyperboloidal}) as we now describe. 

First, applying Eq.~(\ref{modal_KG_eqn_hyperboloidal}) in the limits $\sigma\to \frac{1}{2}^{\pm}$, we obtain
\begin{align}
    J_{\tau\tau} + C_{\tau \sigma} J_{\sigma\tau} + C_{\sigma \sigma} J_{\sigma\sigma} +C_{\tau} J_\tau
    + C_{\sigma} J_\sigma + C_0 J = 0,
    \label{KG_eqn_jumps}
\end{align}
where $J_{\alpha\beta}(\tau) := \left( \phi^>_{,\alpha\beta} - \phi^<_{,\alpha\beta}\right)\Big|_{\sigma=1/2}$, and again 
all $C$ coefficients are evaluated at $\sigma=1/2$. We have $J_{\tau\tau}=d^2 J/d\tau^2=0$, so, recalling also $J=0=J_{\tau}$, we obtain 
\beq
J_{\sigma\sigma} = -(C_{\tau\sigma}\dot{J}_\sigma + C_{\sigma}J_\sigma)/C_{\sigma\sigma}\Big|_{\sigma=1/2} 
\label{Jss},
\eeq
where $\dot{J}_{\sigma}:=dJ_{\sigma}/d\tau=J_{\sigma\tau}$. This is a formula for the jump in the second $\sigma$ derivative, expressed in terms of the jump $J_{\sigma}$ already obtained and its time derivative $\dot{J}_\sigma$ obtainable by differentiating Eq.~(\ref{Js}) with respect to $\tau$.

Jumps in higher $\sigma$ derivatives are constructed recursively in a similar manner, by considering successive $\tau$ derivatives of the field equation (\ref{modal_KG_eqn_hyperboloidal}) evaluated at $\sigma \to \frac{1}{2}^{\pm}$, and noting that $d^nJ/d\tau^n=0$ for all $n\geq 0$. For the jumps in the third and fourth derivatives, we obtain, in this fashion,  
\begin{align}
    J_{\sigma \sigma \sigma} &= - \frac{1}{C_{\sigma \sigma}} \big [ \ddot{J}_{\sigma} + C_{\tau \sigma} \dot{J}_{\sigma \sigma} + J_{\sigma} ( C_0  + C'_{\sigma} ) 
    \nonumber \\
    &\qquad +J_{\sigma \sigma} ( C_{\sigma} + C'_{\sigma \sigma} ) + \dot{J}_{\sigma} ( C_{\tau} + C'_{\tau \sigma} ) \big ] 
\label{Jsss}
\end{align}
and 
\begin{align}
    \label{Jssss}
    J_{\sigma \sigma \sigma \sigma} &= - \frac{1}{C_{\sigma \sigma}} \big [ \ddot{J}_{\sigma \sigma} + ( 2 C'_{\tau \sigma} + C_{\tau} ) \dot{J}_{\sigma \sigma} + C_{\tau \sigma} \dot{J}_{\sigma \sigma \sigma} 
    \nonumber \\
    &\qquad + ( 2 C'_{\sigma \sigma} + C_{\sigma} ) J_{\sigma \sigma \sigma} + ( 2 C'_0 + C''_{\sigma}) J_{\sigma}
    \nonumber \\
    &\qquad + J_{\sigma \sigma} ( C_0 + 2 C'_{\sigma} + C''_{\sigma \sigma} ) 
    \nonumber \\
    &\qquad + ( 2 C'_{\tau} + C''_{\tau \sigma} ) \dot{J}_{\sigma} \big ], 
\end{align}
where all $C$ coefficients and their derivatives are evaluated at $\sigma=1/2$. In each case, the jumps are expressed in terms of lower-order jumps obtained in previous steps, along with their time derivatives, which may be explicitly evaluated in terms of the orbital functions $x_p(\tau)$ and $\varphi_p(\tau)$ and their own time derivatives.

\section{Finite-difference scheme}
\label{sec:fin_diff}

In this work, we develop and test a simple Cauchy evolution code for solving the modal field equation \eqref{modal_KG_eqn_hyperboloidal} with a fixed geodesic source, which can be a scattering geodesic. The code is based on a finite-difference scheme formulated on a fixed mesh in $\tau,\sigma$ coordinates, with uniform sampling intervals $h_{\tau}:=\Delta\tau$ and $h_{\sigma}:=\Delta\sigma$. The scheme we use here is locally fourth-order convergent in the spatial interval $h_\sigma$ (or almost so; see below), and we use a standard fourth-order Runge-Kutta integrator \cite{press2007numerical} to propagate data from one $\tau={\rm const}$ surface to another. The delta-function source is accounted for via suitable jump conditions imposed at $\sigma=1/2$, as detailed below. In preparation for discretization, the (homogeneous part of the) second-order equation \eqref{modal_KG_eqn_hyperboloidal} is first written in a first-order form,
\begin{align}\label{1st_order_form}
    \dot{\phi}&=\Pi,
    \nonumber\\
    \dot{\Pi} &+ C_{\tau \sigma} \Pi' + C_{\tau} \Pi = -C_{\sigma \sigma} \phi''
    - C_{\sigma} \phi' - C_0 \phi ,
\end{align}
where, again, $\ell,m$ indices are dropped for brevity. 

The next step is to prescribe finite-difference formulas for the derivatives $\phi'$, $\phi''$, and $\Pi'$. Each given $\tau = {\rm const}$ slice is divided into $n=1/h_{\sigma}$ equal intervals, where $h_\sigma$ is chosen such that $n$ is an even integer. This defines a vector of coordinate values $\{\sigma_i\}$ with $\sigma_i=i h_{\sigma}$ for $i=0,\ldots,n$, which form a uniform spatial grid on the time slice. We denote the values of the fields $\phi$ and $\Pi$ at $\sigma_i$ by $\phi_i$ and $\Pi_i$, respectively. 

For grid points that are not in the vicinity of the boundaries ($\sigma=0,1$) or the particle ($\sigma =1/2$), we use a simple five-point symmetric stencil that we call our ``vacuum'' (V) finite-difference scheme. Specifically, for $2\leq i\leq (n/2) - 2$ or $(n/2) - 2 \leq i\leq n-2$ we use
\begin{subequations}
\label{vac_FD}
\begin{align}
    (\phi'_i)_\text{V} =& \frac{1}{12 h_{\sigma}} \big [ ( \phi_{i - 2} - \phi_{i + 2} ) - 8 ( \phi_{i - 1} - \phi_{i + 1} ) \big ] 
    \nonumber\\
    &+ \mathcal{O} (h_{\sigma}^4),
    \\
    (\phi''_i)_\text{V} =& \frac{1}{12 h_{\sigma}^2} \big [ - 30 \phi_i + 16 ( \phi_{i - 1} + \phi_{i + 1} ) 
    \nonumber\\
    & \quad- ( \phi_{i - 2} + \phi_{i + 2} ) \big ]  + \mathcal{O} (h_{\sigma}^4).
\end{align}
\end{subequations}
For the boundary points at $\sigma = 0$, we instead use the one-sided stencil
\begin{subequations}
\label{scri_forward_FD}
\begin{align}
    \phi'_0 =& \frac{1}{12 h_{\sigma}} \big ( -25 \phi_0 + 48 \phi_1 - 36 \phi_2 
    + 16 \phi_3 - 3 \phi_4 \big ) 
    \nonumber\\
    &+ \mathcal{O} (h_{\sigma}^4),
    \\
    \phi''_0 =& \frac{1}{12 h_{\sigma}^2} \big ( 45 \phi_0 - 154 \phi_{1} + 214 \phi_{2} - 156 \phi_{3}  
    \nonumber \\
    &\qquad + 61 \phi_{4} - 10 \phi_{5} \big ) + \mathcal{O} (h_{\sigma}^4).
\end{align}
\end{subequations}
The same stencil is used at $\sigma=1$, with the replacements $\phi_i\to\phi_{n-i}$ and $h_\sigma\to-h_\sigma$. 
Since our vacuum stencil uses two neighboring grid points on either side of the evaluation point, we also need to amend the scheme at the next-to-boundary points. For $i=1$, we use the asymmetric stencil
\begin{subequations}
\label{asym_FD_1}
\begin{align}
    \phi'_1 =& \frac{1}{12 h_{\sigma}} \big ( -3 \phi_0 - 10 \phi_1 + 18 \phi_2 - 6 \phi_3 
    \nonumber\\
    &\qquad + \phi_4 \big ) + \mathcal{O} (h_{\sigma}^4),
    \\
    \phi''_1 =& \frac{1}{12 h_{\sigma}^2} \big ( 10 \phi_0 - 15 \phi_1 - 4 \phi_2 + 14 \phi_3  
    \nonumber \\
    &\quad - 6 \phi_4 + \phi_5 \big ) + \mathcal{O} (h_{\sigma}^4),
\end{align}
\end{subequations}
and the same scheme is used for $i=n-1$, with the replacements $\phi_i\to\phi_{n-i}$ and $h_\sigma\to-h_\sigma$. 

To obtain simple derivative stencils that nonetheless retain a high-order convergence even at the particle location, we take advantage of our knowledge of the derivative jumps there, obtained analytically in the previous section. These can be incorporated into our symmetric vacuum scheme so as to correctly account for the jumps between field derivatives on either side of $\sigma=1/2$. A short calculation gives
\begin{subequations}
\label{eq:s0_stencil_P}
\begin{align}
    \phi'_{n/2}  =& (\phi'_{n/2})_\text{V} + \frac{1}{36} \big ( 18 J_{\sigma} 
    - 6 h_{\sigma} J_{\sigma \sigma} + h_{\sigma}^3 J_{\sigma \sigma \sigma \sigma} \big ) 
    \nonumber \\
    & + \mathcal{O} (h_{\sigma}^4),
    \label{eq:s0_phiP_Ds_FD}
    \\
    \phi''_{n/2} =& ( \phi''_{n/2} )_\text{V} + \frac{1}{18 h_{\sigma}} \big ( - 21 J_{\sigma} + 9 h_{\sigma} J_{\sigma \sigma} - 2 h_{\sigma}^2 J_{\sigma \sigma \sigma} \big ) 
    \nonumber \\
    & + \mathcal{O} (h_{\sigma}^3)
    \label{eq:s0_phiP_Ds2_FD},
\end{align}
\end{subequations}
where $(\phi'_{n/2})_{\text{V}}$ is the vacuum scheme from Eq.~\eqref{vac_FD}. Our convention is that these derivative values are the ones obtained by taking one-sided limits $\sigma\to\frac{1}{2}^+$ (recall that both first and second derivatives are generally discontinuous at the particle). 

For next-to-particle points, a similar calculation gives
\begin{subequations}
\label{eq:s0_stencil_Ppm1}
\begin{align}
    \phi'_{n/2 \pm 1}  &= (\phi'_{n/2 \pm 1})_\text{V} + \frac{1}{288} \big ( \mp 24 J_{\sigma} + 12 h_{\sigma} J_{\sigma \sigma} 
    \nonumber \\
    &\qquad \mp 4 h_{\sigma}^2 J_{\sigma \sigma\sigma}
    + h_{\sigma}^3 J_{\sigma \sigma \sigma \sigma} \big ) + \mathcal{O} (h_{\sigma}^4)
    \label{eq:s0_phiPpm1_Ds_FD},
    \\
    \phi''_{n/2 \pm 1} &= ( \phi''_{n/2 \pm 1} )_\text{V} + \frac{1}{288h_\sigma} \big ( 24 J_{\sigma} \mp 12 h_{\sigma} J_{\sigma \sigma} 
    \nonumber \\
    &\qquad + 4 h_{\sigma}^2 J_{\sigma \sigma \sigma}
    \mp h_{\sigma}^3 J_{\sigma \sigma \sigma \sigma} \big ) + \mathcal{O} (h_{\sigma}^3)
    \label{eq:s0_phiPpm1_Ds2_FD},
\end{align}
\end{subequations}
where, as before, our convention is that values at $\sigma=1/2$ are one-sided limits as $\sigma\to\frac{1}{2}^+$.

The stencils for $\Pi'$ are analogous to those for $\phi'$ with the replacements $J_{\sigma} \to \dot{J}_{\sigma}, J_{\sigma \sigma} \to \dot{J}_{\sigma \sigma}, \cdots$

In Eqs.~\eqref{eq:s0_stencil_P} and \eqref{eq:s0_stencil_Ppm1} we have allowed $O(h^3_{\sigma})$ local errors in the second derivative, for simplicity. It is not hard to write down stencils with $O(h^4_{\sigma})$ errors, but these involve either the fifth derivative jump or additional input points beyond the five points used in the vacuum stencil, which somewhat complicates the scheme. As we illustrate in the next section, this slight compromise does not seem to impact the overall quartic convergence of our code empirically.  

With these derivative stencils, Eq.~\eqref{1st_order_form}) becomes a coupled set of first-order ordinary differential equations in $\tau$ for $\phi_i(\tau)$ and $\Pi_i(\tau)$, thought of now as vectors on each time slice. We use (as mentioned) a standard fourth-order Runge-Kutta method (RK4) to evolve these data vectors in time, starting from initial data at some $\tau=\tau_{\rm init}$. When evolving the sourced equation, we set the fields to zero on the initial surface, $\phi_i(\tau_{\rm init})=0=\Pi_i(\tau_{\rm init})$ for all $i$, resulting in a burst of junk radiation, later discarded. This behavior will be demonstrated in our numerical illustrations below.

\section{Code validation: Scalar field}
\label{sec:num_tests}

In this section, we explore the performance of our scalar-field code in a variety of setups. We first consider vacuum evolutions (without a particle source) illustrating the code's long-term stability. We then introduce a particle source, first on a bound orbit and then for scattering, demonstrating the convergence properties of our method and testing its numerical outputs against analytical results and existing data from other codes.

\subsection{Vacuum evolution and stability}
\label{subsec:vacuum_decay}

To consider a vacuum evolution, we set all jumps $J_\sigma,J_{\sigma\sigma},\ldots$ to zero and initiate the field with some nonzero initial data on $\tau=\tau_{\rm init}$, specified below. Our coordinates still carry reference to a fiducial timelike worldline $x^\alpha=y_{p}^\alpha(\tau)$, which for our vacuum tests we choose to specify in two different ways. First, we take (in Schwarzschild coordinates) 
\begin{align}\label{circ}
 y_p^\alpha = (t,R,\pi/2,\Omega t),
\end{align}
with $R={\rm const}=10M$ and $\Omega := (M/R^3)^{1/2}$, corresponding to an equatorial circular geodesic orbit with radius $r=R$. Then we also run with $y_p^\alpha$ corresponding to the scattering geodesic shown in Fig \ref{fig:scatter_orbit},  parametrized by $(\vinf, b) = (0.2, 21M)$. In both cases we start the evolution with a Gaussian initial-data packet, 
\begin{align}
\label{eq:ScalarVacID}
\phi_{\ell m}(\tau_{\rm init},\sigma)=
\exp\left[-\frac{1}{2}\left(\frac{r^*(\tau_{\rm init},\sigma)-10M}{3M}\right)^2\right],
\end{align}
and the initial derivative field $\Pi_{\ell m} (\tau_{\text{init}}, \sigma) \equiv  0$. (We remark that these initial conditions correspond to two physically distinct initial setups in the two cases, so we do not expect the corresponding solutions to agree.)

\begin{figure}[htb]
    \centering
    \includegraphics[width=\linewidth]{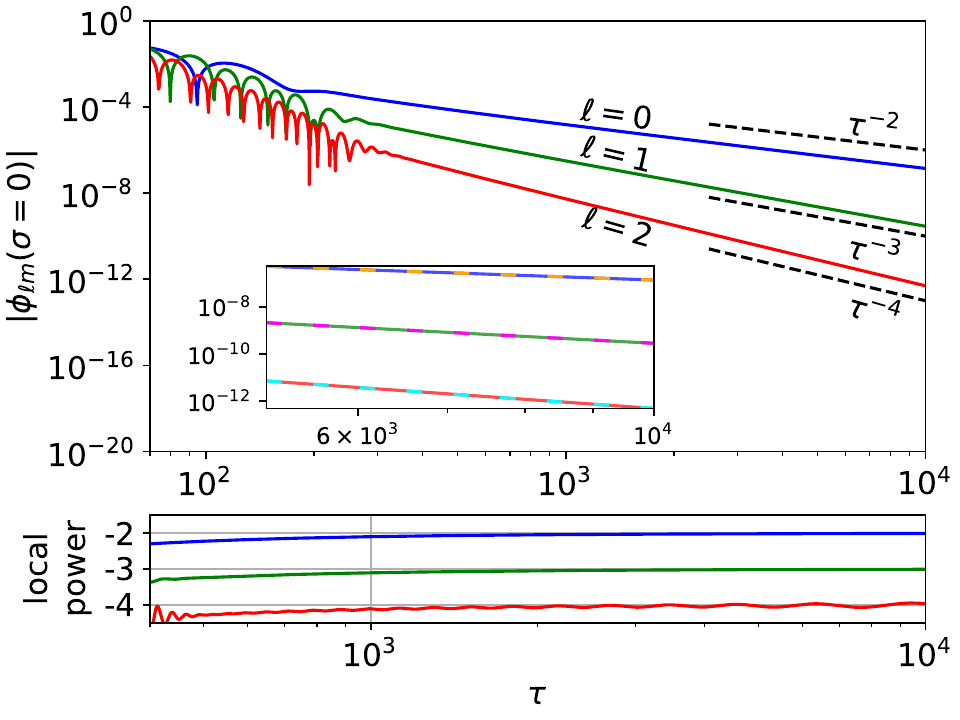}
    \caption{Decay of the $\ell = 0, 1, 2$ modes of the scalar field at $\scri^+$ in the vacuum case, as a function of hyperboloidal time $\tau$. Here, the fiducial worldline in our coordinate definition is a circular geodesic at radius $r = 10M$, and the temporal and spatial finite-difference intervals are $(h_{\tau}, h_{\sigma} ) = (0.016, 0.002)$. The initial data is the narrow Gaussian described in \eqref{eq:ScalarVacID}. Black dashed lines are reference $\tau^{-\ell - 2}$ lines (of arbitrary amplitude) showing the slope of expected power-law tails at late retarded time, with the bottom panel displaying the local power  extracted from the data using $(\tau - \tau_{\rm init}) \dot{\phi} / \phi$. The inset compares with lower-resolution results (dashed lines) obtained with $(h_{\tau}, h_{\sigma} ) = (0.016, 0.004)$, suggesting finite-difference errors in these results are very small. 
    }
    \label{fig:VacCirc_decay}
\end{figure}

\begin{figure}[htb]
    \centering
    \includegraphics[width=\linewidth]{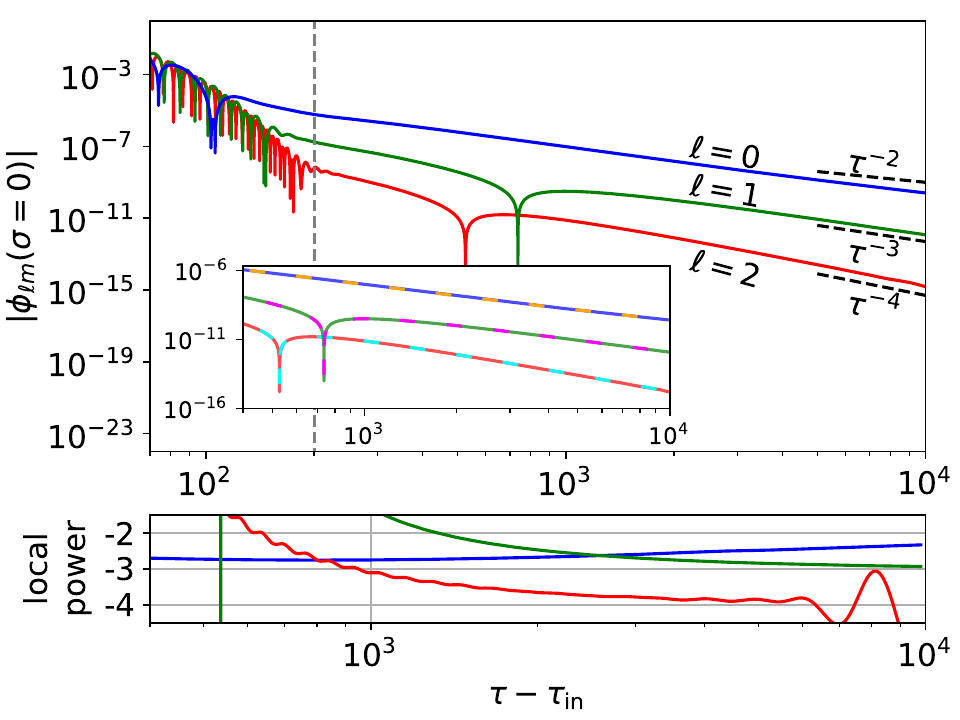}
    \caption{Same as in Fig.~\ref{fig:VacCirc_decay}, but now using a scattering geodesic with $(v_\infty, b) = (0.2, 21M)$ as a fiducial worldline in the coordinates definition. The vertical line marks the periapsis time of the reference geodesic. We took here $(h_{\tau}, h_{\sigma}) = (0.004, 0.001)$, comparing with $(h_{\tau}, h_{\sigma}) = (0.004, 0.002)$ in the inset. The nonphysical wiggly feature at the late-time end of the $\ell=2$ local-power curve is a manifestation of the poorer coordinate performance at large $|\tau|$ discussed in Sec.\ \ref{subsec:scattering}.
    }
    \label{fig:VacScat_decay}
\end{figure}

Figures~\ref{fig:VacCirc_decay} and~\ref{fig:VacScat_decay} show sample results from vacuum evolution runs with circular-orbit-based and scattering-based coordinates, respectively. In both cases, we display the numerical solution at null infinity, $\phi_{\ell m}(\tau,0)$ (which, in this case, is independent of $m$), as a function of hyperboloidal time $\tau$, for $\ell=0,1,2$. We see the expected characteristic ringing behavior at early time, followed by a power-law decay tail at late time, consistent with the expected $\phi \sim \tau^{ -\ell - 2}$ fall-off\footnote{Generic perturbations from compact initial data exhibit the familiar $\phi_{\ell m}\sim u^{-\ell - 2}$ fall-off tail along $\scri^+$ \cite{Price:1972_pert1, Price:1972_pert2}. Using \eqref{trans} we find $u (\sigma \to 0) \propto M (\tau - x_p (\tau)) \sim M \tau (1 - v_{\infty})$ at late time, and thus $\phi_{\ell m} \sim \tau^{-\ell - 2}$ as quoted.}. In the insets we examine the effect of increasing the spatial finite-difference interval $h_\sigma$ by a factor $2$ (at a fixed $h_\tau$), illustrating numerical convergence. Decreasing $h_{\tau}$ by a factor of 2 (at fixed $h_\sigma$) similarly has no visible effect on the data at the display scale of our plot. We will explore the numerical convergence properties of the code more quantitatively in the next subsection.  

Since we are using an explicit (RK4) time integrator, our numerical scheme is limited by the Courant-Friedrichs-Lewy (CFL) condition. To estimate the limiting Courant ratio, we tested the vacuum code for stability for a range of ratios $h_{\tau} / h_{\sigma}$. We find stable evolutions for $h_{\tau} / h_{\sigma} \lesssim 8$ in the case of circular-orbit-based coordinates and $h_{\tau} / h_{\sigma} \lesssim 6$ for our scattering-based coordinates. In the rest of this analysis, we use Courant ratios close to these limits. We then generally find that the finite-difference error in our results comes predominantly from $h_\sigma$, not $h_\tau$.

The main lesson from these vacuum tests is that, with a sufficiently small Courant ratio $h_{\tau} / h_{\sigma}$, our scheme is numerically stable and can be used to evolve the vacuum field equations over a very long physical simulation time. In the examples shown above, we have terminated the simulation after $10^4M$ of physical evolution time, which should suffice for most foreseeable applications involving scattering orbits. At $h_{\sigma}=0.001$ and $h_{\tau}=0.004$, each such simulation took about $140$ minutes of wall time on a single CPU with 500GiB RAM.

\subsection{Circular-orbit source}
\label{subsec:circular}

Next, we consider evolution with a sourcing particle on a circular geodesic orbit, whose worldline is described in Eq.~(\ref{circ}). The jump conditions derived in Sec.~\ref{subsec:jumps} now come into play. Their explicit form becomes much simpler in the circular-orbit case. In particular, we find that \eqref{Js} reduces to
\beq\label{Jsigma_circ}
J_\sigma = -\frac{96\pi M}{R}\sqrt{1-\frac{3M}{R}}\, {Y}^*_{\ell m}(\pi/2,\Omega \tau),
\eeq
while \eqref{Jss} becomes
\begin{align}
J_{\sigma \sigma} &= -4 ( J_{\sigma} + 8 \dot{J}_{\sigma} ),
\nonumber \\
&= -\frac{384 \pi M}{R} \sqrt{1-\frac{3M}{R}}( 8 i m M \Omega - 1 ) Y^*_{\ell m} (\pi/2, \Omega \tau).
\end{align}
From Eq.\ (\ref{Jsss}), we further get
\beq
J_{\sigma \sigma \sigma} = \frac{8}{3} \big [ 504 \ddot{J}_{\sigma} + 198 \dot{J}_{\sigma} + ( 13 + 216 \tilde{V}_{\ell} ) J_{\sigma} \big ]
\eeq
and Eq.\ (\ref{Jssss}) gives
\begin{align}
\label{Jssss_Circ}
J_{\sigma \sigma \sigma \sigma} &= - \frac{64}{3} \big [ 2496 \dddot{J}_{\sigma} + 1944 \ddot{J}_{\sigma} + 2 (187 + 864\tilde{V}_{\ell}) \dot{J}_{\sigma} \big ] 
\nonumber \\
&\quad - \frac{64}{3} J_{\sigma} ( 15 + 648 \tilde{V}_{\ell} + 1296M f (R) \tilde{V}'_{\ell} )
\end{align}
where $\tilde V'_{\ell} (R) = ( d \tilde V_{\ell} / dr )\big|_R$ and where time derivatives can be evaluated via $\partial_\tau=-imM\Omega$. These jumps all have the form ${\rm const}\times e^{-im\Omega t}$.

We note that $m=0$ modes have a static source and hence static physical solutions. We can write these solutions analytically:
\beq
\label{eq:circ_static_sol}
\phi_{\ell 0} = A_\ell  \times \left\{
\begin{array}{ll}
(r/M) P_\ell\left(\frac{r}{M}-1\right)Q_\ell\left(\frac{R}{M}-1\right), & \sigma\geq \frac{1}{2},
\\[1.2ex]
(r/M) P_\ell\left(\frac{R}{M}-1\right)Q_\ell\left(\frac{r}{M}-1\right), & \sigma\leq \frac{1}{2},
\end{array}
\right.
\eeq
where $P_\ell$ and $Q_\ell$ are, respectively, the Legendre polynomial and Legendre function of the second kind\footnote{\label{fn:Qlm_Math}The appropriate function $Q_\ell$ here is the variant with its branch cut from $-\infty$ to $+1$; it is {\it Mathematica}'s `type 3' function, called using $\mathsf{LegendreQ[\ell,0,3,x]}$.}, and 
\begin{equation}
A_\ell = 4\pi\sqrt{1-\frac{3M}{R}}\, {Y}^*_{\ell 0}(\pi/2,0)\ .
\end{equation}

For our sourced runs we set all initial data to zero, $\phi_{\ell m}(\tau_{\rm init}, \sigma) \equiv 0 \equiv \Pi_{\ell m} (\tau_{\rm init}, \sigma)$; the field is then sourced dynamically by the jumps along the particle's worldline. As before, we expect junk radiation from the inconsistent initial data to dissipate rapidly, unveiling the physical (retarded) solution at later times. 

\begin{figure}[h!]
    \centering
    \includegraphics[width=\linewidth]{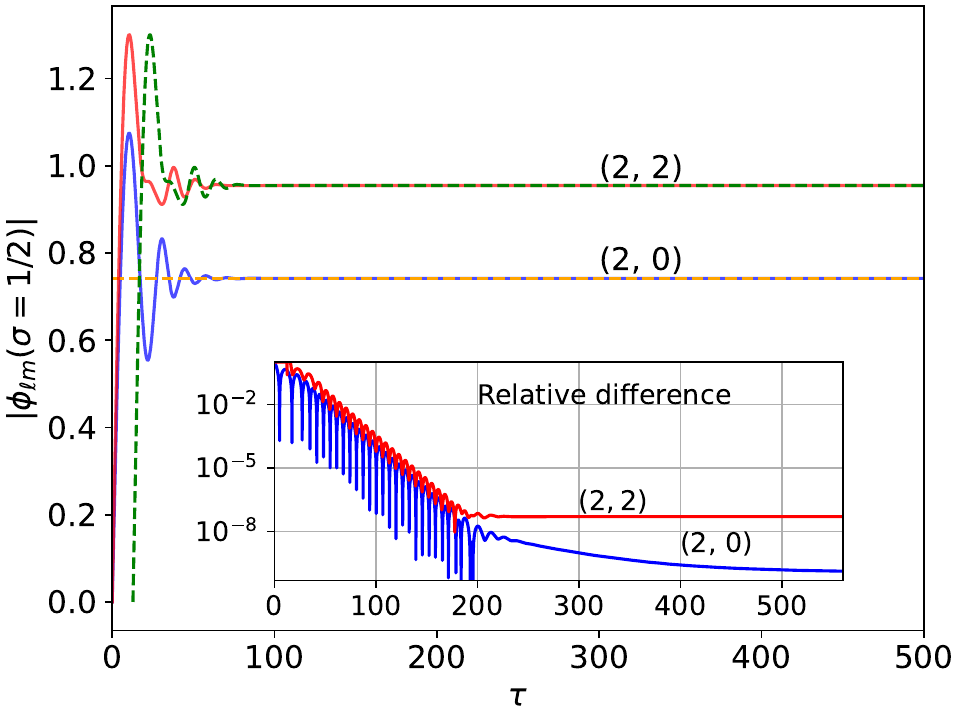}
    \caption{Magnitude of the scalar field along the particle's orbit for the $(2, 0)$ mode (solid blue) and $(2, 2)$ mode (solid red). The particle source is set on a circular geodesic orbit with radius $r = 10M$. The orange dashed line indicates the value of the static solution \eqref{eq:circ_static_sol} and the green dashed line shows the numerical value obtained from the uv code of Ref.~\cite{Long:2021TDmetric}. The inset shows the relative difference between our hyperboloidal solution and the reference result in each case. While the figure is truncated at $\tau = 500$, we have evolved the system stably up to $\tau = 5000$.}
    \label{fig:circ_abs_psiP_tau}
\end{figure}

This behavior is manifest in Fig.~\ref{fig:circ_abs_psiP_tau}, showing the values of $|\phi_{20}|$ and $|\phi_{22}|$ as functions of $\tau$ along the particle's worldline ($\sigma=1/2$). We see the initial burst of junk radiation, which then decays rapidly, leaving behind a stationary solution, as expected. For $\phi_{2 0}$, comparison against the analytical formula (\ref{eq:circ_static_sol}) shows a very good agreement, with a difference---presumably due to residual junk radiation---that decays over time and is as small as $\sim 10^{-10}$ fractionally by the end of our simulation.  For $\phi_{22}$, we instead compare with numerical results from the double-null time-domain code of Ref.~\cite{Long:2021TDmetric} (hereafter ``uv code''), also demonstrating a good agreement. Our further analysis suggests that the residual difference here is dominated by error from the uv code.

\begin{figure}[h!]
    \centering
    \includegraphics[width=\linewidth]{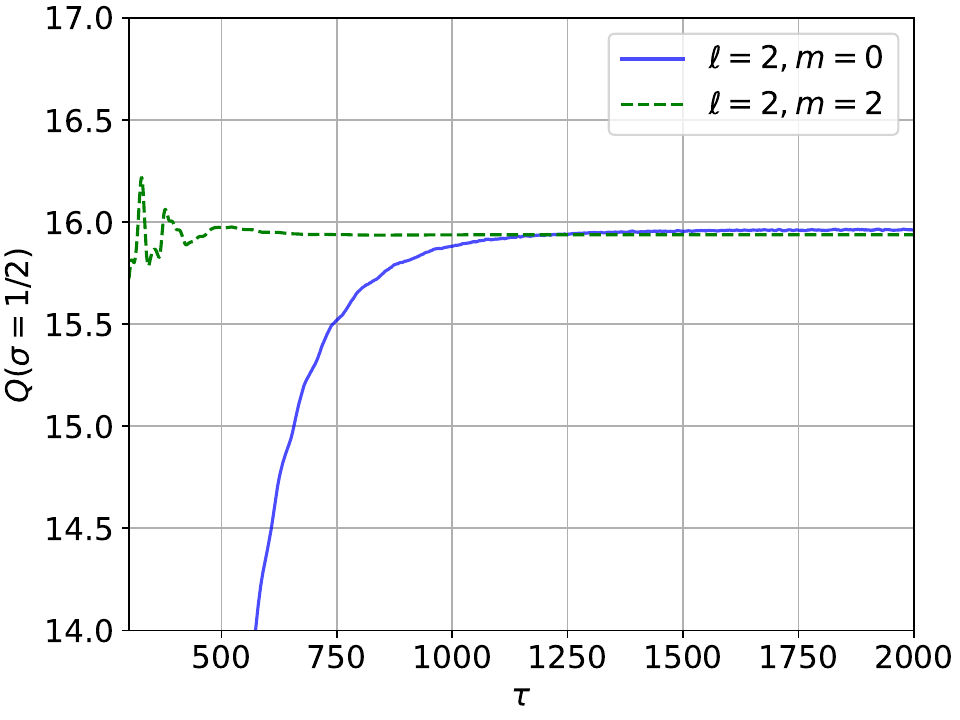}
    \caption{The $\sigma$-convergence rate $Q$ of the $(2, 0)$ and $(2, 2)$ modes for a particle in circular orbit of radius $R = 10M$ is shown along the particle's worldline. For the static mode, the convergence diagnostic~\eqref{eq:Qsig_stat} is used with 2 spatial resolutions $h_{\sigma} = 0.002, 0.001$. For the $(2, 2)$ mode, we use the convergence index~\eqref{eq:Qsig} with three spatial resolutions $h_{\sigma} = 0.004, 0.002, 0.001$. In both cases, the temporal resolution is held fixed at $h_{\tau} = 0.008$. Once initial junk subsides, the convergence appears to be quartic (fourth order) in both cases. 
    }
    \label{fig:circ_QsigP_tau}
\end{figure}

Figure \ref{fig:circ_QsigP_tau} explores the convergence rate of our finite-difference scheme with respect to the spatial interval $h_\sigma$ (at fixed $h_\tau$). For the $(2,0)$ mode we use as a diagnostic the convergence index 
\begin{align}
\label{eq:Qsig_stat}
Q=\left|\frac{\phi[h_{\sigma}]-\phi_\text{exact}}{\phi[h_{\sigma}/2]-\phi_\text{exact}}\right|,
\end{align}
where arguments in square brackets indicate the resolution applied, and $\phi_\text{exact}$ is the exact analytical solution from Eq.~(\ref{eq:circ_static_sol}).
For the $(2,2)$ mode, in the absence of an analytical solution, we instead use 
\begin{align}
\label{eq:Qsig}
Q=\left| \frac{\phi[h_{\sigma}]-\phi[h_{\sigma}/2]}{\phi[h_{\sigma}/2]-\phi[h_{\sigma}/4]} \right|. 
\end{align}
In both cases, a quartic convergence (to the correct value, in the former case) should yield $Q\sim 16$. In Fig.~\ref{fig:circ_QsigP_tau} we plot $Q$ along the particle's orbit for the two modes, confirming that, once junk radiation subsides, the convergence is approximately quartic. 

We have also tested the convergence of our RK4 evolution scheme, varying $h_\tau$ at fixed $h_\sigma$. We typically find that, for values of $h_\tau$ and $h_\sigma$ satisfying the Courant condition, the error of our RK4 scheme falls below machine precision. 

\subsection{Scattering source}
\label{subsec:scattering}

Next we implement the scattering-orbit source from Fig.\ \ref{fig:scatter_orbit}. Here, we cannot leverage any simplifications when computing the jumps, and must instead evaluate them in full using Eqs.~\eqref{Js} and \eqref{Jss}--\eqref{Jssss}. As in the circular-orbit case, we start with zero initial data, letting the field be sourced by the imposed jumps on the scattering orbit, and later discarding initial junk radiation.  

\begin{figure}[h!]
    \centering
    \includegraphics[width=\linewidth]{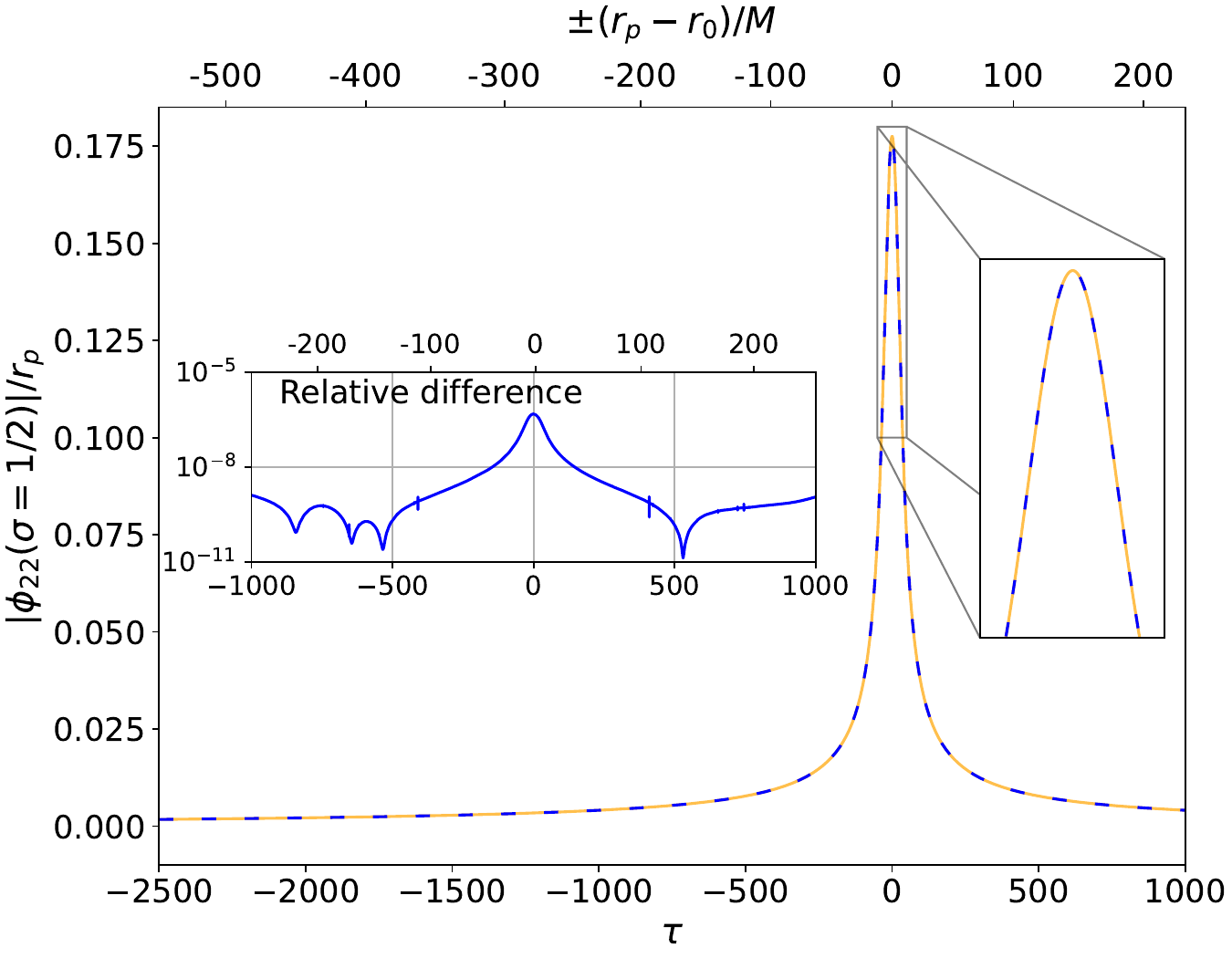}
    \caption{Magnitude of $\phi_{22}/r$ along the worldline of a scalar charge on a scattering orbit with $(v_\infty, b) = (0.2, 21M)$ (see Fig.~\ref{fig:scatter_orbit}). The lower x-axis shows the hyperboloidal time $\tau$ along the worldline, while on the upper x-axis we show the radial distance from periapsis, $\pm(r_p - r_{0})/M$, with negative (positive) values corresponding to the incoming (outgoing) leg of the scattering geodesic. The main plot shows the results of the hyperboloidal scheme (solid orange), with grid spacings $h_{\tau} = 0.004$ and $h_{\sigma} = 0.001$, compared with results from the uv scheme (dashed blue). In both cases, the particle begins at an initial separation of $r_p = 1200M$ on the incoming leg, and the initial junk-dominated interval is discarded. The simulation ends at a separation of $r_p = 250M$ on the outgoing leg. The inset on the right expands the periapsis region for clarity. The inset on the left shows the relative difference between the two curves from the main plot. 
    }
    \label{fig:scat_22_field_res}
\end{figure}

\begin{figure}[h!]
    \centering
    \includegraphics[width=\linewidth]{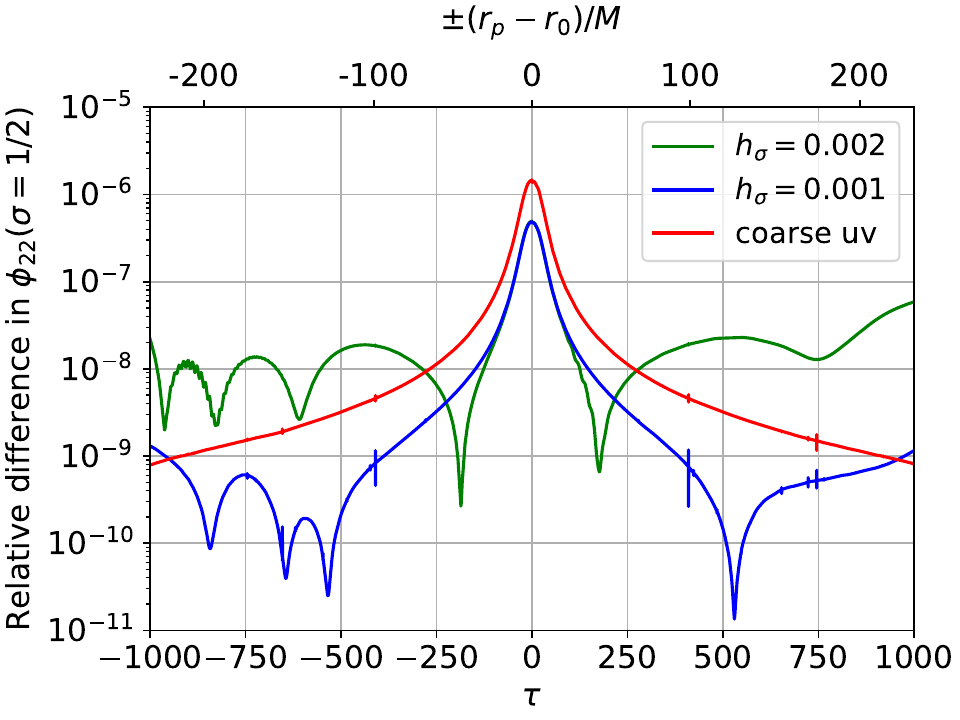}
    \caption{Relative difference between the hyperboloidal and uv-code $\phi_{22}$ results from Fig.~\ref{fig:scat_22_field_res} with varying numerical resolution. The blue curve is a reproduction of the relative difference shown in the inset of Fig.~\ref{fig:scat_22_field_res}. The green curve shows the relative difference when the hyperboloidal code is run at a (twice) lower spatial resolution. A comparison suggests that the error near periapsis, where both curves agree, comes predominantly from the uv code; however, the error from the hyperboloidal code appears to dominate at larger radii. This conclusion is corroborated by the red curve, showing the relative difference between two sets of uv data: the one from Fig.~\ref{fig:scat_22_field_res} and another obtained with a (twice) coarser uv grid. }
    \label{fig:uv_hyp_comp}
\end{figure}

Figure \ref{fig:scat_22_field_res} shows our numerical results for the $(2,2)$ mode of the field recorded along the scattering geodesic, as compared to the numerical solution from the uv code of Refs.~\cite{Long:2021TDmetric} and \cite{barack_self-force_2022}. After the initial junk subsides, we see a very good agreement between the two sets of results. The relative difference (invisible to the eye on the scale of the main plot) is shown in the inset to be mere $\sim 10^{-6}$ near the periapsis, and smaller still further out along the orbit. To better understand how the accuracy of our results varies along the orbit, in Fig.~\ref{fig:uv_hyp_comp}, we compare with lower-resolution results from both hyperboloidal and uv codes. The comparison suggests that in the strong-field region near periapsis, the difference is dominated by error from the uv code, i.e.~the hyperboloidal results are more accurate there. However, the situation is reversed at large $r_p$, where the performance of the hyperboloidal code appears to deteriorate faster than that of the uv code. 

The poorer performance of our hyperboloidal coordinates at large $r_p$ is explained as follows.  Near $\mathscr{I}^+$, i.e., for $\sigma\ll 1$, the coordinate relation (\ref{r*_trans}) becomes $r_*/M\simeq x_p(\tau)+1/\sigma^{2}$. At sufficiently small $|\tau|$, this produces the desired compactification relation $r_*\sim M/\sigma^2$ on each $\tau={\rm const}$ slice. However, for $|\tau|$ large enough that $x_p(\tau)\gtrsim 1/h_\sigma^2$, this relation is effectively lost, since the $x_p(\tau)$ term dominates over the $1/\sigma^{2}$ term even at the grid points nearest to $\mathscr{I}^+$. As $|\tau|$ increases, the grid points become ever more concentrated around the particle's location, at the expense of resolution near $\mathscr{I}^+$. In the theoretical limit $|\tau|\to\infty$, the entire array of grid points on the $\tau={\rm const}$ slice would shrink to a single point at $r_*/M=x_p(\tau)$. This loss of resolution near $\mathscr{I}^+$ for large $r_p$ becomes less severe with increasing spatial resolution (smaller $h_{\sigma}$), but could cause a problem in practice. A possible mitigation could be based on a readjustment of our hyperboloidal coordinates, modifying the compactification power $1/\sigma^2\to 1/\sigma^{n>2}$ in Eqs.~(\ref{H}) and (\ref{K}). Further mitigation strategies can involve the use of nonuniform and/or adaptive spatial grids. Such remedies could be applied as needed in applications where the issue proves problematic.

\begin{figure}[h!]
    \centering
    \includegraphics[width=\linewidth]{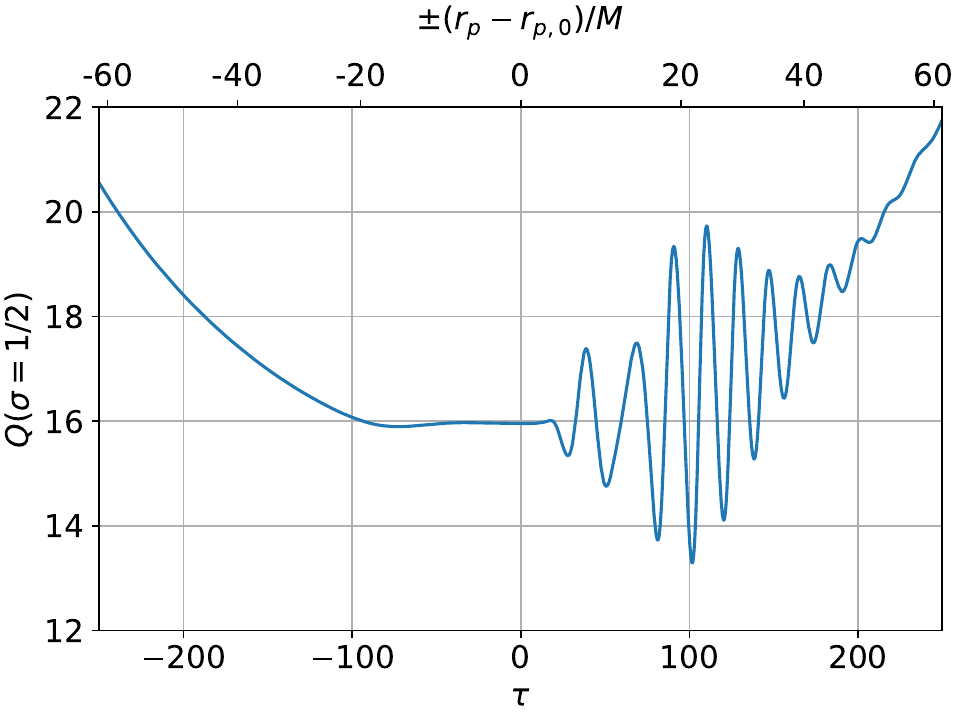}
\textit{}    \caption{The $\sigma$-convergence rate $Q$ from~\eqref{eq:Qsig} for the $(2, 2)$ mode of the scalar field along the scattering worldline at fixed $h_{\tau} = 2 \times 10^{-3}$. For this test, we have used the three spatial resolutions $h_{\sigma} = 2 \times 10^{-3}, 1 \times 10^{-3}$ and $5 \times 10^{-4}$. $Q\sim 16$ indicates quartic (fourth-order) numerical convergence.}
    \label{fig:scat_Qsig_rp}
\end{figure}

In Fig.~\ref{fig:scat_Qsig_rp} we attempt to quantify the numerical convergence rate in the scattering case, again using the diagnostic $Q$ from Eq.\ (\ref{eq:Qsig}). We see an approximate quartic convergence near the periapsis, as in the circular-orbit case. However, we also see a distinct oscillatory feature in $Q$ post-periapsis in the outgoing leg, and the convergence rate becomes less clear further out along the orbit on both legs.  We suspect the oscillations are related to the familiar phenomenon of post-periapsis ringing \cite{Thornburg:2019ukt, Nasipak:2019hxh, barack_self-force_2022} (attributed to quasinormal mode excitations at periapsis passage), which is too small here to be visibly manifest in the field itself. The behavior at large radii is presumably a consequence of the overall poorer coordinate performance discussed above.

\section{Application to the gravitational Hertz potential}
\label{sec:hertz_Teuk}

Up until this point, we have been developing and testing our new time-domain evolution scheme in the familiar setting of a scalar-field model. This enabled a study of our method's performance in an environment where comparison results are readily available for benchmarking. We remind the reader, however, that the main motivation for our new method is the need to address the problem of uncontrolled growing modes that plague naive characteristic schemes for time evolution of the $|s|=2$ Teukolsky equation \cite{Long:2021TDmetric}. We now finally turn to the question of whether the new method indeed overcomes this problem. 

In this section we apply our method to the Teukolsky equation for the $s=-2$ gravitational Hertz potential used for radiation-gauge metric reconstruction in self-force applications \cite{Pound:2013faa}. Our goal is to demonstrate that the method remains numerically stable, and that, crucially, the problematic growing modes are dynamically suppressed. We shall demonstrate this here in the relatively simple example of circular geodesic orbit sources, where some (analytical and numerical) comparison results are available.  The case of scattering, our ultimate goal, requires further development (notably the derivation of suitable jump conditions) and will be presented in follow-up work. 

\subsection{$s=-2$ Teukolsky equation in hyperboloidal coordinates}

We consider a particle moving around a Schwarzschild black hole on a circular geodesic orbit $y_p(\tau)$ as in Eq.~(\ref{circ}). Instead of a scalar charge, the particle now carries a mass $\mu\ll M$, which sources a gravitational perturbation. In one approach to constructing this perturbation, one first derives a certain Hertz potential $\Phi^+$ as a particular solution to the sourced $s=-2$ Teukolsky equation. From that potential (and certain ``completion'' pieces \cite{Merlin:2016boc}), one reconstructs the physical metric perturbation in a so-called no-string incoming radiation gauge, using a procedure developed in Ref.~\cite{Pound:2013faa}. This approach has been a cornerstone of frequency-domain self-force calculations for bound orbits. 

For our current time-domain study, we follow Ref.~\cite{Long:2021TDmetric} in decomposing $\Phi^+$ into multipole modes using
\begin{align}
\label{eq:Hertz_decomposition}
\Phi^+ = r^3 f^2 \sum_{\ell = 2}^\infty \sum_{m = -\ell}^\ell \phi_{\ell m}^+(u,v) {}_{-2}Y_{\ell m}(\theta, \varphi),
\end{align}
where ${}_{-2}Y_{\ell m}$ are spin-weighted spherical harmonics of spin weight $-2$, and where $u,v$ are the standard Eddington-Finkelstein coordinates introduced earlier. Off of the particle's worldline, the time-radial functions $\phi^+(u,v)$ (with their $\ell m$ indices hereafter suppressed for brevity) satisfy the modal vacuum $s=-2$ Teukolsky equation \cite{Long:2021TDmetric}
\begin{align}
    \phi^+_{,uv} + U(r) \phi^+_{,u} + V(r) \phi^+_{,v} + W(r) \phi^{+} = 0
\label{eq:BPT_uv},
\end{align}
where 
\begin{subequations}
\begin{align}
    U (r) &=  \frac{2M}{r^2}, \\
    V (r) &= -\frac{2 f}{r}, \\
    W (r) &= \frac{f}{4} \bigg ( \frac{(\ell -1) (\ell +2)}{r^2} - \frac{2 M }{r^3} \bigg ).
\end{align}
\label{eq:BPT_UVW}
\end{subequations}

To write Eq.~(\ref{eq:BPT_uv}) in terms of our comoving hyperboloidal coordinates $(\tau,\sigma)$, we need the Jacobian derivatives 
\begin{subequations}
\label{eq:tau_sig_1deriv}
\begin{align}
    2M\tau_{,u} &= 1 + H' \xi \zeta_+,
    \\
    2M\sigma_{,u} &= - \xi \zeta_+,
    \\
    2 M \tau_{, v} &= 1 - \xi H' \zeta_-,
    \\
    2 M \sigma_{, v} &= \xi \zeta_-,
\end{align}
\end{subequations}
where $\xi \coloneqq (K' + \alpha' x_p - H' \alpha \dot{x}_p)^{-1}$ and $\zeta_{\pm} \coloneqq 1 \pm \alpha \dot{x}_p$, along with
\begin{subequations}
\label{eq:tau_sig_2deriv}
\begin{align}
    4 M^2 \tau_{, u v} &= \xi H' ( \zeta'_+ \sigma_{, v} + \dot{\zeta}_+ \tau_{, v}) + \zeta_+ \big [ \xi H'' \sigma_{, v}  
    \nonumber \\
    &\quad + H' ( \xi' \sigma_{, v} + \dot{\xi} \tau_{, v} ) \big ],
    \\
    4 M^2 \sigma_{, u v} &= - \xi (\zeta'_+ \sigma_{, v} + \dot{\zeta}_+ \tau_{, v}) - \zeta_+ (\xi' \sigma_{, v} + \dot{\xi} \tau_{, v}).
\end{align}
\end{subequations}
 
In terms of these derivatives, we rewrite the modal Teukolsky equation (\ref{eq:BPT_uv}) in the form
\begin{align}
\label{eq:BPT_hyper_vac}
    {\phi}^+_{,\tau\tau} + C^{+}_{\tau \sigma} \phi^+_{,\tau\sigma} + C^{+}_{\sigma \sigma} \phi^+_{,\sigma\sigma} + C^{+}_{\tau} \phi^+_{,\tau} 
    &+ C^{+}_{\sigma} \phi^+_{,\sigma} \nonumber\\
    &+ C^{+}_0 \phi^+ = 0,
\end{align}
where the coefficients are
\begin{subequations}
\label{eq:BPT_hyper_coeff}
\begin{align}
    C^+_{\tau \sigma} &= \frac{\sigma_{, v}}{\tau_{, v}} + \frac{\sigma_{, u}}{\tau_{, u}}, \\
    C^{+}_{\sigma \sigma} &= \frac{\sigma_{, u} \sigma_{, v}}{\tau_{, u} \tau_{, v}}, \\
    C^{+}_{\tau} &= \frac{\tau_{, u v} + U \tau_{, u} + V \tau_{, v}}{\tau_{, u} \tau_{, v}}, \\
    C^{+}_{\sigma} &= \frac{\sigma_{, u v} + U \sigma_{, u} + V \sigma_{, v}}{\tau_{, u} \tau_{, v}}, \\
    C^{+}_0 &= \frac{W}{\tau_{, u} \tau_{, v}},
\end{align}
\end{subequations}
with the potentials $U(r)$, $V(r)$, and $W(r)$ from Eqs.~(\ref{eq:BPT_UVW}) now considered functions of $\tau,\sigma$.  

For our numerical implementation, we also require the jumps in $\phi^+$ and its $\sigma$ derivatives along the particle's worldline. We shall denote these by $J^+$, $J^+_\sigma$, etc., mirroring the notation used in the scalar-field case. The source of the Hertz potential $\phi^+$ in Eq.~(\ref{eq:BPT_hyper_vac}) is not directly available, but in Ref.~\cite{Long:2021TDmetric} (building on preliminary results in \cite{Barack:2017oir}), two of us have derived the jump conditions that it must satisfy, and these jumps suffice for our purposes. The general form of these jumps is rather complicated, and involves a numerical integration. However, in the case of a circular orbit, the jumps simplify considerably and can be expressed algebraically. Since the expressions remain rather unwieldy, we delegate them to the Appendix. Importantly, and unlike in the scalar-field case, the Hertz potential is {\it discontinuous} across the worldline, with $J^+\ne 0$ in general. The explicit expressions for $J^+$, as well as for $J^+_\sigma$ and $J^+_{\sigma\sigma}$, are given in Appendix \ref{app:hertz_jumps}. 

Usefully, as in the scalar-field case, an explicit analytical solution can be written for the axisymmetric, $m=0$ modes of the Hertz potential (which are time-independent for a circular-orbit source), providing a convenient comparison in this case. Citing from Eq.~(85) of Ref.~\cite{Barack:2017oir}, the static, axisymmetric solution is given by
\begin{align}
\label{eq:hertz_static}
\phi^+_{\ell 0} = \begin{cases}
    C^>_{\ell} (R) \phi^{\text{P}}_{\ell} (r), &\quad \sigma > 1/2, \\
    C^<_{\ell} (R) \phi^{\text{Q}}_{\ell} (r), &\quad \sigma \leq 1/2,
    \end{cases}
\end{align}
where $R$ is the radius of the circular orbit, and $\phi^{\text{P}}_{\ell}$, $\phi^{\text{Q}}_{\ell}$ are the two linearly independent solutions of the static Teukolsky equation,
\begin{subequations}
\begin{align}
    \phi^{\text{P}}_{\ell} &= \frac{P_{\ell 2} (x)}{\sqrt{\lambda (\lambda + 2)} (r - 2 M)},
    \\
    \phi^{\text{Q}}_{\ell} &= \frac{Q_{\ell 2} (x)}{\sqrt{\lambda (\lambda + 2)} (r - 2 M)},
\end{align}
\end{subequations}
with $P_{\ell m}$ and $Q_{\ell m}$ being the associated Legendre functions of the first and second kinds, respectively (see Footnote~\ref{fn:Qlm_Math} for a clarification regarding the appropriate version of $Q_{\ell m}$ to be used). Here, we have defined $x := (r - M)/M$ and $\lambda := (\ell + 2) (\ell - 1)$, and the coefficients $C^>$ and $C^<$ in Eq.~\eqref{eq:hertz_static} are given by
\begin{subequations}
\begin{align}
    C^>_{\ell} (R) &= \left.\frac{R^4 f^3}{M} \left( -J^+ \frac{d\phi_\ell^{\text{Q}}}{dr} + J^+_r \phi^{\text{Q}}_\ell \right)\right|_{r=R},
    \\
    C^<_{\ell} (R) &= \left.\frac{R^4 f^3}{M} \left ( -J^+ \frac{d\phi^{\text{P}}_\ell}{dr} +J^+_r \phi^{\text{P}}_\ell \right)\right|_{r=R}.
\end{align}
\end{subequations}

No such analytical solutions are known for $m\ne 0$ modes. Instead we can compare with numerical results obtained by two of us in Ref.\ \cite{Long:2021TDmetric}, which were constructed from time-domain numerical solutions of the Regge-Wheeler equations.

\subsection{Finite-difference scheme}

We use a slight adaptation of the finite-difference method presented in Sec.~\ref{sec:fin_diff} to evolve Eq.\ (\ref{eq:BPT_hyper_vac}) from ``zero'' initial conditions, $\phi^+(\tau_{\rm init},\sigma)\equiv 0 \equiv \Pi^+(\tau_{\rm init},\sigma)$, with the analytically prescribed jump conditions from Appendix \ref{app:hertz_jumps}. At vacuum points away from the particle we again calculate spatial derivatives using the stencil in Eq.~\eqref{vac_FD}, and we use~\eqref{scri_forward_FD} and~\eqref{asym_FD_1} for boundary and next-to-boundary points. 

However, the treatment of particle and near-particle points is now different, since we need to take account of the nonzero jump $J^+$ in the field itself. The appropriate generalization of our scalar-field stencils in Eq.~\eqref{eq:s0_stencil_P} is
\begin{subequations}
\label{eq:phiP_FD}
\begin{align}
    (\phi^+)'_{n/2} &= \phi'_{n/2} - \frac{21}{36 h_{\sigma}} J^+ + \mathcal{O} (h_{\sigma}^4)
    \label{eq:phiP_Ds_FD},
    \\
    (\phi^+)''_{n/2} &= \phi''_{n/2} + \frac{5}{4 h_{\sigma}^2} J^+ + \mathcal{O} (h_{\sigma}^3),
    \label{eq:phiP_Ds2_FD}
\end{align}
\end{subequations}
where $\phi'_{n/2}$ and $\phi''_{n/2}$ represent the scalar-field stencils. Similarly, Eq.~\eqref{eq:s0_stencil_Ppm1} for the next-to-particle points generalizes as
\begin{subequations}
\label{eq:phiPm1_FD}
\begin{align}
    (\phi^+)'_{n/2 - 1} &= \phi'_{n/2 - 1} - \frac{7}{12 h_{\sigma}} J^+
     + \mathcal{O} (h_{\sigma}^4)
    \label{eq:phiPm1_Ds_FD},
    \\
    (\phi^+)''_{n/2 - 1} &= \phi''_{n/2 - 1} - \frac{5}{4 h_{\sigma}^2} J^+  + \mathcal{O} (h_{\sigma}^3)
    \label{eq:phiPm1_Ds2_FD},
\end{align}
\end{subequations}
and
\begin{subequations}
\label{eq:phiPp1_FD}
\begin{align}
    (\phi^+)'_{n/2 + 1} &= \phi'_{n/2 + 1}  + \frac{1}{12 h_{\sigma}} J^+  + \mathcal{O} (h_{\sigma}^4)
    \label{eq:phiPp1_Ds_FD},
    \\
    (\phi^+)''_{n/2 + 1} &= (\phi''_{n/2 + 1})^{\text{V}} - \frac{1}{12 h_{\sigma}^2} J^+  + \mathcal{O} (h_{\sigma}^3)
    \label{eq:phiPp1_Ds2_FD}.
\end{align}
\end{subequations}
Now that the field is discontinuous at the particle, we need to be clear about the meaning of $\phi^+_n$. Our convention is that this value represents the ``right-hand'' limit $\lim_{\sigma\to(1/2)^+}\phi_n^+$, so the ``left-hand'' limit is now given by $\phi^+_n-J^+$. Since in our scheme the stencil at $i = n/2 - 2$ uses the left-hand limit of the field point at $n/2$, the stencil at that point must be amended too:
\begin{subequations}
\label{eq:phiPm2_FD}
\begin{align}
    (\phi^+)'_{n/2 - 2} &= \phi'_{n/2 - 2}  + \frac{J^+}{12 h_{\sigma}} + \mathcal{O} (h_{\sigma}^4),
    \\
    (\phi^+)''_{n/2 - 2} &= \phi''_{n/2 - 2}  + \frac{J^+}{12 h_{\sigma}^2} + \mathcal{O} (h_{\sigma}^4).
\end{align}
\end{subequations}
This completes the description of our finite-difference scheme for the $s=-2$ Hertz potential. 

As they stand, the above stencils require as input the jumps $J^+, J^+_{\sigma},\ldots ,J^+_{\sigma \sigma\sigma\sigma}$ along with their time derivatives. For our test implementation here, we wish to avoid having to calculate jumps in the third and fourth $\sigma$ derivatives, which are complicated. Setting those to zero in Eqs.~(\ref{eq:phiP_FD})--(\ref{eq:phiPm2_FD}) yields a finite-difference scheme that is still quadratically convergent (as demonstrated below), which should suffice for our purpose here, namely probing the stability of our implementation for $s=-2$. In what follows, we thus adopt a simpler version of Eqs.~(\ref{eq:phiP_FD})--(\ref{eq:phiPm2_FD}), with $J^+_{\sigma \sigma\sigma}\equiv 0 \equiv J^+_{\sigma \sigma\sigma\sigma}$.

\subsection{Numerical results: Hertz potential}
\label{subsec:hertz_test}

\begin{figure}
    \centering
    \includegraphics[width=\linewidth]{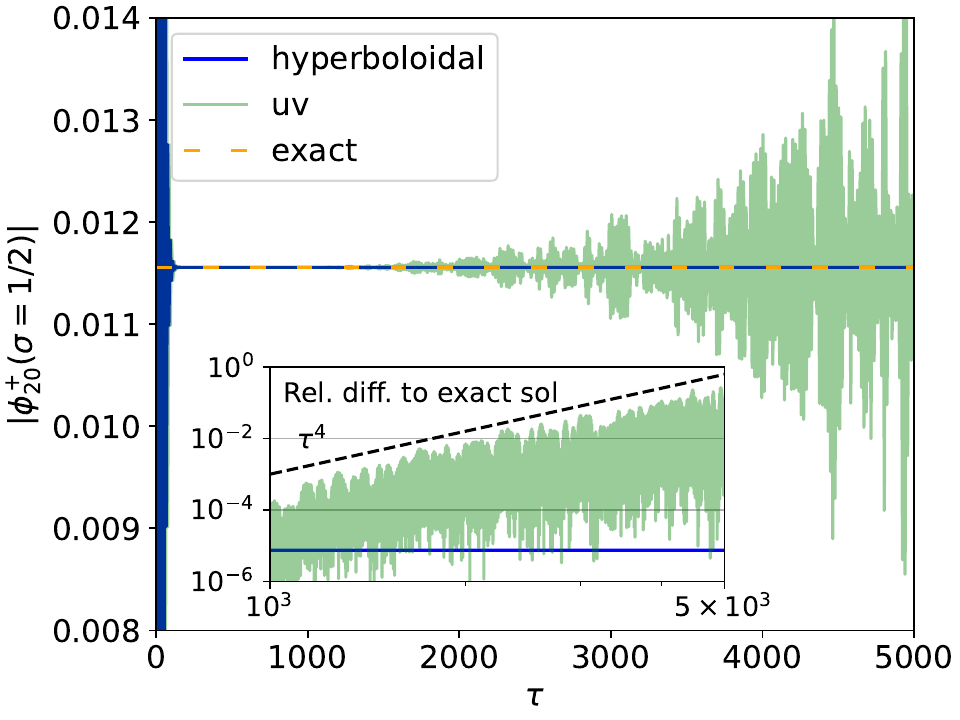}
    \caption{Long-term stability of the hyperboloidal scheme (in blue) in evolution of the $(2, 0)$ mode of the Hertz potential with spin-weight $s = -2$. The source is on a circular geodesic of radius $R \approx 7.12 M$ corresponding to $x_p = 9$, and the field is plotted along the circular orbit as a function of time $\tau$. We have used $h_{\tau} = 0.008$ and $h_{\sigma} = 0.001$. In comparison are results from a direct characteristic evolution using the uv code of Ref.~\cite{Long:2021TDmetric} (in green), and the analytical solution from~\eqref{eq:hertz_static} (in dashed orange). The inset shows the relative difference between each of the numerical datasets and the analytical solution, from which it is clear that the hyperboloidal scheme (blue) is immune to the growing mode problem, while the uv results (green) exhibit the $\tau^4$ growth.}
    \label{fig:Hertz_circ20}
\end{figure}

\begin{figure}
    \centering
    \includegraphics[width=\linewidth]{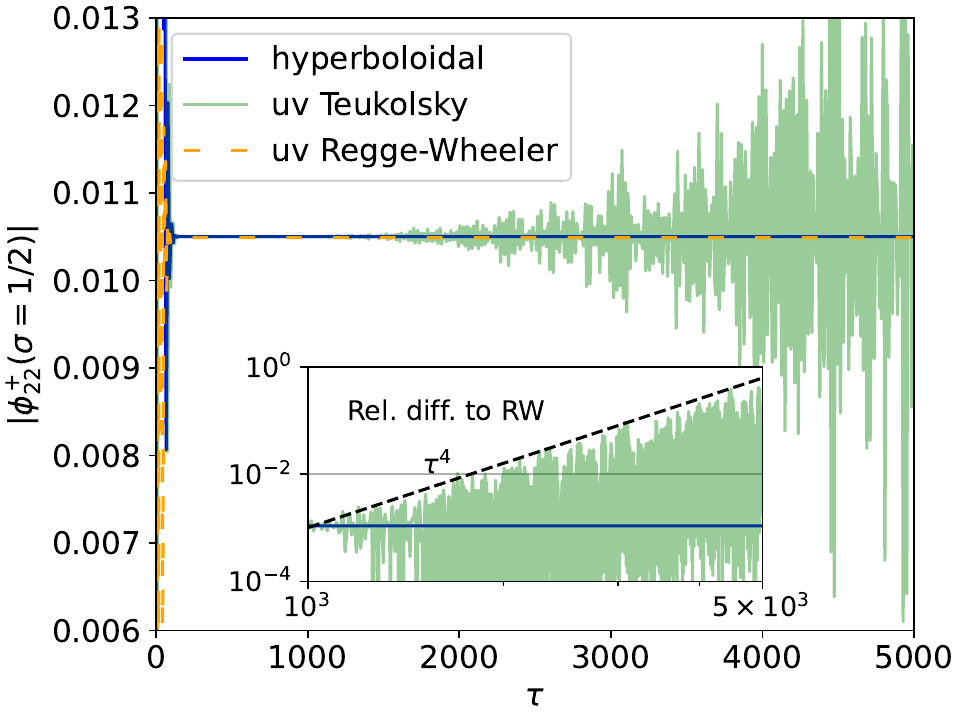}
    \caption{Same as Fig.~\ref{fig:Hertz_circ20}, here for the $(2, 2)$ mode of the Hertz potential. In the absence of an analytical solution for benchmarking, we instead use here (dashed orange) numerical solutions obtained from differentiation of the Regge-Wheeler variable following the method of Ref.~\cite{Long:2021TDmetric}.}
    \label{fig:Hertz_circ22}
\end{figure}

Figures~\ref{fig:Hertz_circ20} and~\ref{fig:Hertz_circ22} show numerical results for the Hertz potential modes $\phi^+_{20}$ and $\phi^+_{22}$, respectively, from evolutions with a circular-orbit source. Plotted are the fields as functions of $\tau$ along the particle's worldline. Shown superposed are results from the uv code of Ref.~\cite{Long:2021TDmetric} for the same setup. We see that, in the hyperboloidal scheme, the evolution proceeds stably over $5000M$ of physical time, long beyond the point at which the uv scheme fails due to the uncontrolled growth of $\sim t^4$ solution modes. No sign of this behavior is detected in our hyperboloidal scheme. Furthermore, the hyperboloidal results are seen to agree well with the reference solutions, i.e.~the analytical solution (\ref{eq:hertz_static}) for the $(2, 0)$ mode, and the Hertz potential derived from the Regge-Wheeler solution \textit{à la} Ref.~\cite{Long:2021TDmetric} for the $(2, 2)$ mode. We have further run our code for a variety of other mode numbers and radii $R$, and out to $\tau=10^4M$, with the troublesome $\sim t^4$ behavior never showing itself. 

\begin{figure}
    \centering
    \includegraphics[width=\linewidth]{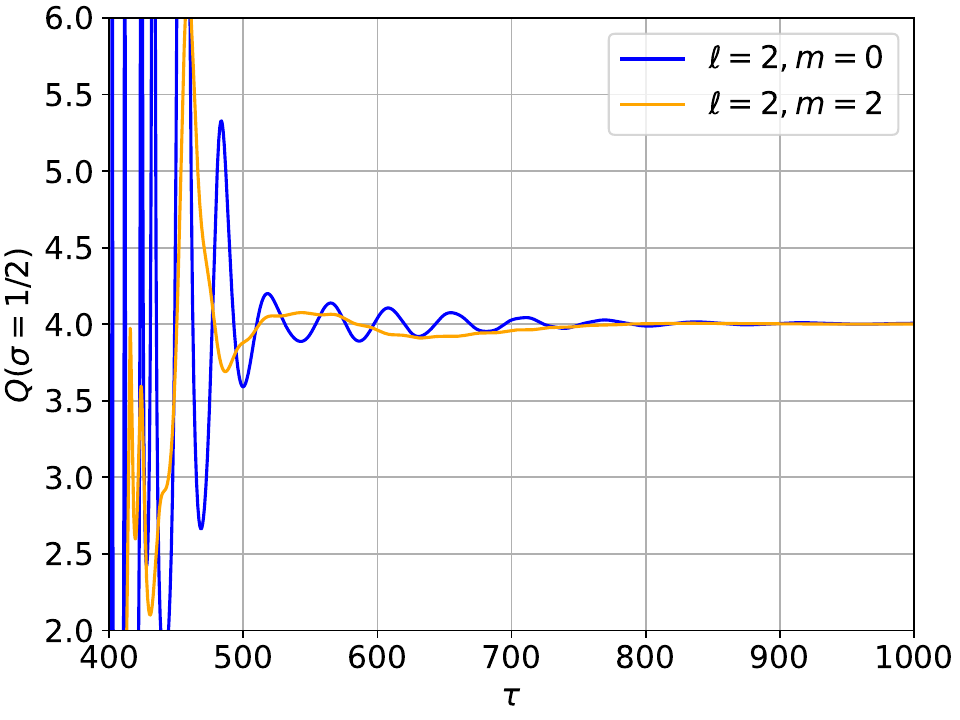}
    \caption{The $\sigma$-convergence rate $Q$ along the particle's worldline for the $(2, 0)$ mode (dashed blue) and $(2, 2)$ mode (solid orange) of the Hertz potential. For the axisymmetric mode, we use the diagnostic~\eqref{eq:Qsig_stat} with the exact solution given by~\eqref{eq:hertz_static} and two spatial resolutions $h_{\sigma} = 0.002, 0.001$ and time step $h_{\tau} = 0.008$. For the $(2, 2)$ mode, we use the diagnostic~\eqref{eq:Qsig} with same time step and an extra spatial resolution $h_{\sigma} = 0.004$. In both cases, we see that once initial junk subsides we recover the expected quadratic convergence.}
    \label{fig:Hertz_Qsig}
\end{figure}

In Fig.~\ref{fig:Hertz_Qsig} we examine the convergence rate of our Teukolsky code. As expected, the convergence appears to be quadratic once transients from initial junk radiation subside.

The message to take from the above experiments is that the use of compactified spatial coordinates appears efficient in suppressing the $\sim t^4$ growing mode in time-domain simulations of the $s=-2$ Teukolsky equation. This confirms the expectation expressed in Ref.~\cite{Long:2021TDmetric}.

\section{Conclusions}
\label{sec:discussion}

The problem of growing modes in time-domain numerical integration of the $s=\pm 2$ Teukolsky equation \cite{Long:2021TDmetric} has been a roadblock in the program to calculate the gravitational self-force in extreme-mass-ratio scattering. In this work, we have confirmed the expectation, expressed in Ref.~\cite{Long:2021TDmetric}, that the problem can be effectively mitigated through the use of spatially compactified hyperboloidal coordinates. This we have illustrated here with an explicit numerical implementation for the $s=-2$ Hertz potential in the case of circular geodesic motion.  The road to full gravitational self-force calculations via metric reconstruction from curvature scalars is now open.

Specializing to a Schwarzschild background, in Sec.~\ref{sec:hyperboloidal_comoving}, we presented a set of compactified hyperboloidal coordinates that are ``comoving'' with respect to a fixed timelike geodesic worldline and discussed their various properties. The {\it comoving} property is particularly advantageous, as it dramatically eases the imposition of jump conditions along the worldline in the 1+1-dimensional treatment of the sourced field equations. In Sec.~\ref{sec:scalar_field_hyper}, we applied this method to the Klein-Gordon equation on a Schwarzschild background. We wrote down the (mode-decomposed, 1+1-dimensional) scalar-field equations in terms of our new coordinates, and described the derivation of jump conditions along the reference worldline in these coordinates. In Secs.~\ref{sec:fin_diff} and \ref{sec:num_tests}, we then developed and implemented a new finite-difference code for evolving the modal equations from initial data, with the appropriate jump conditions associated with a sourcing scalar charge moving along the reference worldline. We tested our code in both scenarios of circular and scattering orbits, illustrating the method's stability, numerical convergence, and accuracy. 

In Sec.~\ref{sec:hertz_Teuk}, we then applied our method to the $s=-2$ Teukolsky equation for the Hertz potential that can be used to generate metric perturbations for a sourcing mass particle. We demonstrated that, at least in the case of circular motion, our code evolves stably over many dozens of orbital cycles, and returns accurate results. This we have contrasted with numerical solutions from evolution in null coordinates without compactification, which develop a nonphysical, noisy feature whose amplitude grows as $\sim t^{4}$ at late time, rendering the results unusable. 

In this work, we have stopped short of applying our method to the Hertz potential in a scattering scenario, our ultimate goal. We are hoping to present that in follow-up work, starting with the somewhat nontrivial task of  deriving the appropriate jump conditions along the particle's worldline in terms of our comoving hyperboloidal coordinates. Once we have a numerical scheme for the Hertz potential, the computation of the metric perturbation and self-force along the scattering orbit is a straightforward task, following the procedure summarized in Ref.~\cite{Long:2021TDmetric}. (However, how to control the asymptotic reference frames in which scattering ``invariants'' are computed from the self-force remains an open problem that would need to be addressed separately.)

Our comoving hyperboloidal scheme provides a general framework for time-domain calculations with a reference worldline with perceivable applications that are not limited to self-force calculations. Of course, our proposed coordinates are not unique, and neither are they even most natural. For instance, a choice that is more ``symmetric'' with respect to the particle would require the height function to be stationary at the worldline, i.e.~$H'(1/2)=0$. Additionally, we could require local symmetry of the shift vector at the particle by demanding $K''(1/2)=0$. We have experimented with a range of such aesthetics-improving choices, but none proved as numerically stable as the one ultimately adopted here. Of course, there exist completely different choices of source-adapted hyperboloidal coordinates that may be explored; see Ref.~\cite{Zenginoglu:2025_BridgingTimes} for some alternatives.

A potential weakness of our scheme was identified in the discussion at the end of Sec.~\ref{sec:num_tests}, namely, the pathological behavior of our coordinates in the large $r_p$ limit. This may be particularly troublesome 
in the scattering scenario, where the orbital radius is unbounded. We have listed possible mitigations to be investigated in such scenarios, including tweaking our coordinates to the effect of improving sampling resolution near $\scri^+$. A more systematic solution to the problem, in the scattering case, would be to formulate exact, analytical Cauchy initial data for the scattering problem at $\tau\to-\infty$, achievable in principle using a perturbative weak-field analysis. This should allow us to begin the scattering evolution at smaller orbital radii, and at the same time suppress the amplitude of initial junk radiation. Work to derive such initial conditions is underway, with promising preliminary results~\cite{IC}.     
Finally, we comment that our compactified-coordinates scheme should {\it in principle} be able to tackle also the $s=+2$ evolution problem, although the exponential growth of the problematic mode in that case can pose additional difficulties in practice. We intend to study this in future work.

\begin{acknowledgments}

A.V. would like to thank An{\i}l Zengino{\u{g}}lu for helpful discussions. A.V. further acknowledges the support of the European Consortium for Astroparticle Theory in the form of an Exchange Travel Grant.
L.B. acknowledges support from the STFC via Grant No. ST/B001170/1.
O.L. is funded by the European Union (ERC grant GWSky/101167314). Views and opinions expressed are however those of the author(s) only and do not necessarily reflect those of the European Union or the European Research Council Executive Agency. Neither the European Union nor the granting authority can be held responsible for them.
R.P.M. acknowledges support from the Villum Investigator program supported by the VILLUM Foundation (grant no.\ VIL37766) and the DNRF Chair program (grant no.\ DNRF162) by the Danish National Research Foundation. The Center of Gravity is a Center of Excellence funded by the Danish National Research Foundation under grant No. DNRF184. 
A.V., O.L. and R.P.M. would like to thank the Erwin Schrödinger International Institute for Mathematics and Physics (ESI) for the opportunity to present this work at the workshop ``Hyperboloidal Foliations and their Application''.  
This work makes use of the Black Hole Perturbation Toolkit~\cite{BHPToolkit}, specifically the \texttt{pybhpt}~\cite{Nasipak_pybhpt_2026,Nasipak:2025tby}, \texttt{SpinWeightedSpheroidalHarmonics}~\cite{wardell_2026_22809903}, and \texttt{KerrGeodesics}~\cite{niels_warburton_2023_8108265} packages.

\end{acknowledgments}
\section*{Data Availability}

There are no publicly available research data or software supporting this manuscript. Requests for further information or data should be sent to the authors.


\appendix
\section{Jump conditions for the Hertz potential modes $\phi^+_{\ell m}$}
\label{app:hertz_jumps}

Here we describe the jump conditions that the Hertz potential modes $\phi^+_{\ell m}(\tau,\sigma)$ and their derivatives must satisfy along timelike circular geodesic orbits outside a Schwarzschild black hole. These are needed for our numerical implementation in Sec.~\ref{sec:hertz_Teuk}. The jump conditions were first derived in Ref.~\cite{Barack:2017oir}, and here we only cite the results and reformulate them in terms of our $\tau,\sigma$ coordinates. 

In Ref.~\cite{Barack:2017oir} it was shown how the jumps in $\phi^+_{\ell m}$ are obtained from the jumps in the modes of the Weyl scalar $\Psi_0$, themselves deduced from the source of the $s=2$ Teukolsky equation. Writing 
\begin{align}
    \Psi_0 = (r^5 f^2)^{-1} \sum_{\ell = 2}^{\infty} \sum_{m = -\ell}^{\ell} \psi_{\ell m} (t, r) {}_2 Y_{\ell m} (\theta, \varphi),
\end{align}
where ${}_2 Y_{\ell m}$ are spin-2 weighted spherical harmonics, Ref.~\cite{Barack:2017oir} obtained, for a circular geodesic of radius $r=R$,
\begin{align}
    [\psi_{\ell m}] &= 8 \pi \mu \gamma_R R^2 \big [ (y^2 - i \tilde{m} \Omega R) \mathcal{Y} (t)
     - i f_R R\, \Omega \mathcal{Y}_{\theta} (t) \big ]
\end{align}
and 
\begin{align}
    [\psi_{\ell m,r}] &= 4 \pi \mu \gamma_R R \big \{ \big [ 2 y^2 - 2 - y (\lambda - 4) 
    \nonumber \\
    &\quad + \tilde{m}^2 (1 + y^2 \gamma_R^4)/f_R - 2 i \tilde{m} R \Omega (3 - 7y) /f_R \big ] \mathcal{Y}(t) 
    \nonumber \\
    &\quad + 2 (-3 i f_R R \Omega + \tilde{m} ) \mathcal{Y}_{\theta} (t) + f_R \mathcal{Y}_{\theta \theta} (t) \big \}.
\end{align}
Here we use the notation 
\begin{align}
    [\psi(t)]=\lim_{\epsilon\to 0}[\psi(t,R+\epsilon)-\psi(t,R-\epsilon)],
\end{align}
and introduce
\begin{subequations}
\begin{align}
    \gamma_R = (1 - 3M/R)^{-1/2}, &\qquad y = M/R, 
    \\
    f_R=1-2M/R, & \qquad \Omega=(M/R^3)^{1/2},
    \\
    \lambda = (\ell + 2) (\ell - 1), &\qquad \tilde{m} = m/\gamma_R^2,
\end{align}
\end{subequations}
as well as $\mathcal{Y}^{\ell} = {}_2 Y^*_{\ell 0} (\pi/2, 0)$, 
with $\mathcal{Y}^{\ell}_{\theta}$ and $\mathcal{Y}^{\ell}_{\theta \theta}$ representing, respectively, the first and second derivatives of ${}_2 Y^*_{\ell 0} (\theta, 0)$ with respect to $\theta$, evaluated at $\theta=\pi/2$. The jump in the $t$ derivative is obtained via 
\begin{align}
    [\psi_{\ell m,t}] = -im\Omega [\psi_{\ell m}],
\end{align}
and from this, the jump in the derivative with respect to advanced time coordinate $v=t+r_*$ (needed below) is constructed as
\begin{align}
    [\psi_{\ell m,v}] = \frac{f_R}{2} [\psi_{\ell m,r}] + \frac{1}{2} [\psi_{\ell m,t}].
\end{align}

Given these jumps in the Weyl scalar modes, Ref.\ \cite{Barack:2017oir} showed that the jumps in the Hertz potential can be constructed via  
\begin{align}
    [\phi^+_{\ell m}] &= \frac{(-1)^m}{\tilde{\Delta}} \big ( \hat d [\psi^*_{\ell, -m}] - \hat b [(\psi^*_{\ell, -m})_{,v}] \big ),
    \\
    [\phi^+_{\ell m, v}] &= \frac{(-1)^m}{\tilde{\Delta}} \big (\hat a [(\psi^*_{\ell, -m})_{,v}] - \hat c [\psi^*_{\ell, -m}] \big ),
\end{align}
where, recall, an asterisk denotes complex conjugation, and where the coefficients are
\begin{subequations}
\begin{align}
    \tilde{\Delta} &= \frac{1}{4} f_R^4 R^8 \big [ \lambda^2 (\lambda + 2)^2 + (12 m M \Omega)^2 \big ],
    \\
    \hat a &= \frac{1}{2} f_R^2 \lambda ( \lambda + 2 ) R^4 - 2 f_R m^2 R^6 \Omega^2 (\lambda + 6 y) 
    \nonumber \\
    &\quad + 16 i m^3 M R^6 \Omega^3 - 2 i m R^5 \Omega \big [ \lambda -2 (\lambda +5) y^2
    \nonumber \\
    &\quad - ( \lambda - 3 ) y \big ],
    \\
    \hat b &= 8 i m^3 R^8 \Omega^3 - 4 i m R^6 \Omega  \big [ \lambda - 6 y^2 - 2 (\lambda - 1) y \big ],
    \\
    \hat c &= f_R^2 \lambda (\lambda + 2) R^3 (1 - y) + 2 i f_R m^3 R^6 \Omega ^3 (\lambda + 22 y) 
    \nonumber \\
    &\quad - i f_R m R^4 \Omega \big [ \lambda (\lambda + 5) + 12 y^3 - 2 (4 \lambda + 17) y^2 
    \nonumber \\
    &\quad - 2 \left( \lambda^2 + 2 \lambda - 6 \right ) y \big ] + 16 m^4 M R^6 \Omega^4  
    \nonumber \\
    &\quad - 2 m^2 R^5 \Omega^2 \big [ 3 \lambda + 24 y^3 + 2 (\lambda - 23) y^2 
    \nonumber \\
    &\quad + (15 - 7 \lambda ) y \big ]
    \\
    \hat d &= \frac{1}{2} f_R^2 \lambda  (\lambda +2) R^4+16 i f_R m^3 R^7 \Omega ^3 
    \nonumber \\
    &\quad - 2 i f_R m R^5 \Omega \big [ 3 \lambda - 12 y^2 - 5 (\lambda - 1) y \big ] 
   \nonumber \\
   &\quad + 8 m^4 R^8 \Omega^4 - 2 m^2 R^6 \Omega^2 \big ( 3 \lambda - 24 y^2 
    \nonumber \\
   &\quad + 2 (5 - 3 \lambda ) y \big ).
\end{align}
\end{subequations}
The jumps in the derivatives with respect to $t$ and $r_*$ are then (henceforth suppressing $\ell,m$ indices for brevity)
\begin{align}
    [\phi^+_{,t}]&=-im\Omega [\phi^+], \\
    [\phi^+_{,r_*}]&=2[\phi^+_{,v}]+im\Omega [\phi^+].
\end{align}
Jumps in high-order derivatives can be obtained iteratively by applying the modal Teukolsky equation (\ref{eq:BPT_uv}) repeatedly, as explained in \cite{Barack:2017oir}.

For our implementation in Sec.~\ref{sec:hertz_Teuk} we instead need the jumps in $\tau,\sigma$ derivatives. These are easily obtained using the chain rule with the coordinate transformation equations (\ref{trans}). Caution is required about the sign convention: recall that in the text we define the jump in the Hertz potential modes $\phi_{\ell m}^+(\tau,\sigma)$ via
\begin{align}
    J^+(\tau)=\lim_{\epsilon\to 0}[\phi^+(\tau,1/2+\epsilon)-\phi^+(\tau,1/2-\epsilon)],
\end{align}
and that $\sigma$ is a monotonically {\em decreasing} function of $r$. Therefore,
\begin{align}
    J^+=-[\phi^+].
\end{align}
For the jumps in the first and second $\sigma$ derivatives we find, being mindful about this sign conversion and using the chain rule, we find
\begin{subequations}
\begin{align}
    J^+_{\sigma} &= - M H' [\phi^+_{, t}] - M (\alpha' R_* + K') [\phi^+_{, r^*}],
    \\
    J^+_{\sigma \sigma} &= - M^2 H'^2 [\phi^+_{, t t}] - 2 M^2 H' (\alpha' R_* + K') [\phi^+_{, t r^*}] 
    \nonumber \\
    &\quad - M H'' [\phi^+_{, t}] - M^2 (\alpha' R_* + K')^2 [\phi^+_{, r^* r^*}]
    \nonumber \\
    &\qquad - M (\alpha'' R_* + K'') [\phi^+_{, r^*}],
\end{align}
\end{subequations}
where the functions $H, K, \alpha$, and their derivatives are all evaluated at $\sigma = 1/2$, and $R_*$ is the constant value of $x_p$ for the circular geodesic. Jumps in $\tau$ derivatives are computed using $\partial_\tau\to -imM\Omega$; for instance, $\dot{J}^+ = -i m M \Omega J^+$.


\bibliographystyle{apsrev4-1.bst}
\bibliography{bibitems}

\end{document}